\documentclass[aps,prd,reprint,nofootinbib]{revtex4-2}
\usepackage{graphicx}
\usepackage{amsmath,amssymb}
\usepackage[T1]{fontenc}
\usepackage{lmodern}
\usepackage{microtype}
\usepackage{hyperref}
\hypersetup{hidelinks,pdftitle={Dark Matter Model Implications of a Candidate Sub-halo Near the Sun, and the Data Needed to Discriminate Among Them},pdfauthor={Stefano Profumo and Sukanya Chakrabarti}}
\newcommand{\msun}{M_\odot}
\newcommand{\kms}{\mathrm{km\,s^{-1}}}
\newcommand{\sigmam}{\sigma/m_\chi}
\begin{document}
\title{Dark Matter Model Implications of a Candidate Sub-halo Near the Sun, and the Data Needed to Discriminate Among Them}
\author{Stefano Profumo}
\affiliation{Department of Physics, University of California, Santa Cruz, CA 95064}
\author{Sukanya Chakrabarti}
\affiliation{Department of Physics \& Astronomy, University of Alabama in Huntsville, Huntsville, AL 35899}
\date{\today}

\begin{abstract}
Pulsar-timing accelerations point to a localized mass concentration near the Sun, consistent with a dark matter sub-halo. We work out what such an object implies for the nature of dark matter, organizing the implications by their observable effect: suppression or enhancement of structure formation, internal structure and survival after formation, and emission or scattering by the constituents. We discuss ultralight wave dark matter, thermal and nonthermal warm dark matter, dark acoustic oscillations and dark matter--baryon scattering, elastic and inelastic self-interactions, invisible decays with recoil, primordial black holes, indirect detection, and a possible connection to the LUX-ZEPLIN high-energy recoil candidate. We then ask what new data would discriminate among these models. The existing accelerations are measured outside the tidal radius, where they are blind to the interior mass distribution, to the object's velocity, and to the sub-halo population. We identify four routes that lift these degeneracies: accelerations measured inside the halo, which separate point-mass, cuspy, and cored profiles; lensing and accretion, which test whether the mass is compact; the morphology of annihilation or decay emission, which probes the inner profile; and a search for a comoving stellar population, which tests whether the object is dark and measures its velocity. We also contrast this program with pulsar-timing-array searches for transiting dark matter substructure, which probe vastly lighter objects through a different observable.
\end{abstract}
\maketitle

\section{Introduction}
\label{sec:intro}
Galactic accelerations inferred from pulsar timing provide a gravitational route to dark substructure without requiring a luminous tracer bound to the object. Binary orbital-period derivatives, after subtracting intrinsic and apparent-motion contributions, constrain line-of-sight accelerations. Spatially correlated residuals relative to a Galactic potential model can then motivate a localized perturber. Chakrabarti et al.\ report a candidate near the Sun with Bayes factors of approximately $20$--$40$ and characteristic mass of a few $10^7\msun$ \cite{Chakrabarti2025}. A later analysis by Wang et al.\ finds no significant residual signal under its assumptions and a low predicted detection probability for the current binary sample \cite{Wang2026}. The present paper therefore asks what such an object \emph{would} imply if its nature and structure were established.

The inference has several distinct levels. A localized gravitational perturbation need not be an extended, self-bound dark matter halo. Establishing an extended halo would still not determine its inner cusp, core size, constituent identity, or infall mass. The reported NFW model fixes $r_s=0.1$ kpc and $c=30$. These values were motivated by the sizes of sub-halos of comparable mass in cosmological simulations \cite{ErkalBelokurov2015,Moline2017}, because the available data could not constrain them (Sec.~\ref{sec:reference-params} shows that $r_s$ remains unconstrained when it is freed); a successful fit with that template is therefore not a measurement of those quantities \cite{Chakrabarti2025,Zhu2025}. All profile-based calculations below are conditioned on this distinction.

We organize the microphysics by its observable effect: primordial suppression or enhancement of structure; subsequent relaxation, heating, cooling, and disruption; phase-space and profile signatures; and non-gravitational signals. This includes PBHs and mixed dark sectors as well as elementary-particle models. The same object can test several mechanisms, but the underlying profile and selection uncertainties are shared and must not be counted as independent evidence.

The paper has two parts. Sections~\ref{sec:reference}--\ref{sec:lz} state what the current observations would imply for particle models, conditional on the candidate's confirmation; Section~\ref{sec:future} then asks which new measurements would discriminate among those models: accelerations measured inside the halo, lensing and accretion as tests of compactness, the morphology of any annihilation or decay signal, and a search for bound stars; Section~\ref{sec:synthesis} summarizes both parts and lays out the calculations they open.

For binary pulsars, the line-of-sight (LOS) Galactic acceleration \cite{Chakrabarti2021} is measured from the observed drift rate of the binary period, $P_{b}$:

\begin{equation}
\dot{P}_{b}^{\rm Obs} = \dot{P}_{b}^{\rm Gal} + \dot{P}_{b}^{\rm Shk} + \dot{P}_{b}^{\rm GR},
\label{eq:pdot}
\end{equation}

correcting for the Shklovskii term that arises from the transverse motion of the binary, $\dot{P}_{b}^{\rm Shk}$, and the orbital decay from the general-relativistic gravitational radiation emission, $\dot{P}_{b}^{\rm GR}$. Here $\dot{P}_{b}^{\rm Shk} = \mu^{2} d P_{b}/c$, and $\dot{P}_{b}^{\rm GR}$ is a function of the masses and eccentricities of the binary \cite{Weisberg2016}. The LOS Galactic acceleration is then $a_{\rm Gal} = \dot{P}_{b}^{\rm Gal} c / P_{b}$. Uncertainties in the distance (via parallax), proper motion, and masses and eccentricities are incorporated within an MCMC framework. Subsequent work generalized this approach to solitary pulsars, using an empirical magnetic-braking calibration \cite{Donlon_SC_2025}, and including its intrinsic scatter as an additional uncertainty. The inference of the dark matter sub-halo in Ref.~\cite{Chakrabarti2025} rests on identifying a localized residual in the LOS acceleration relative to a smooth Galactic potential:
\begin{equation}
\Delta a(\mathbf{r}_i) = a_{\rm los,measured}(\mathbf{r}_i) - a_{\rm Gal}^{\rm mod}(\mathbf{r}_i).
\label{eq:residual}
\end{equation}
Here $a_{\rm los,measured}(\mathbf{r}_i)$ is the measured LOS Galactic acceleration of pulsar $i$ at position $\mathbf{r}_i$, and $a_{\rm Gal}^{\rm mod}(\mathbf{r}_i)$ is the LOS acceleration predicted at the same position by a model Galactic potential. Reference~\cite{Chakrabarti2025} considered several such potentials, including an isothermal disk and an exponential disk, each with and without a fixed spiral perturbation. It employed 27 binary pulsars (351 possible pairs), required a pair separation of at most 0.5 kpc, and retained pairs whose residual acceleration exceeds the maximum expected from the baryonic components, $6\times10^{-10}~\mathrm{cm\,s^{-2}}$. This threshold acts on the mean absolute residual of the pair and carries no significance requirement, so for most of the five pairs that pass it, it is crossed by central values consistent with zero; only J1640+2224/J1713+0747 has a mean $S/N>3$ in the derived line-of-sight acceleration. Around this pair, Ref.~\cite{Chakrabarti2025} then added binary pulsars with $S/N>0.5$ and spin-derived accelerations with $S/N>1$, and fit the sub-halo and Galactic-potential parameters simultaneously, with a position prior restricted to within 0.5 kpc in each Cartesian coordinate of the pair midpoint and a mass prior uniform in $\log_{10}M_{\rm sub}$ over $10^5$--$10^9\msun$. Appendix~\ref{app:obs2} gives the full specification of the search and fit, including the places where the published text of the Letter differs from the analysis as performed.

For the Galactic potential models and sub-halo parameterizations considered, the fit yields Bayes factors of $\sim20$--40. These are not Bayes factors for a point null. Each is the ratio of the prior probability (1/4) to the posterior probability, estimated as a sample fraction, of the low-mass region $M_{\rm sub}<10^6\msun$, i.e., an interval analogue of the Savage--Dickey density ratio, so its value depends on the mass and position priors. The $Z$ posterior lies against the edge of the position box, so the position uncertainties and the Bayes factors are conditional on that choice; a larger box lowers the significance of the mass, reflecting the degeneracy between sub-halo position and mass that accelerations alone leave \cite{Chakrabarti2025}. Using the public v3 catalog \cite{Donlon_SC_2025} with $R_\odot=8.122$ kpc \cite{GRAVITY2018} and $Z_\odot=0$, we find that 9 of the 351 binary pairs have separation $\leq0.5$ kpc (one of them the two components of the double pulsar), that all five pairs of Table~1 of Ref.~\cite{Chakrabarti2025} are among the nine, and that J1640+2224/J1713+0747 (mean $S/N=3.47$) is the only one with mean $S/N>3$. If the $S/N>3$ requirement is regarded as part of the search definition, the number of effective search sites is therefore one rather than five, and the five-volume estimate below should be read in that light.

The Appendix of Ref.~\cite{Chakrabarti2025} gives an analytic estimate of approximately \(5\%\) for the prior-predictive probability of encountering a sub-halo of $10^7M_\odot$ in five randomly distributed search volumes, assuming a local sub-halo fraction \(f_{\rm sub}=0.03\); it is not the outcome of a null simulation of the pair search, and no look-elsewhere correction over the 351 pairs is reported. It also adopts $\rho_{\rm DM}=5\times10^8\msun\,\mathrm{kpc^{-3}}$, which does not correspond to the quoted $0.3~\mathrm{GeV\,cm^{-3}}$ (equal to $7.9\times10^6\msun\,\mathrm{kpc^{-3}}$; $5\times10^8\msun\,\mathrm{kpc^{-3}}$ is $19~\mathrm{GeV\,cm^{-3}}$). With the corrected density, the same equations give a per-volume probability of $1.6\times10^{-4}$ and a five-volume probability of $8\times10^{-4}$ for $f_{\rm sub}=0.03$, the original values being reproduced with the original density. The equations as written, however, omit the factor $4\pi/3$ of the spherical search volume. Including it raises both probabilities by a factor of $4.19$, to $6.7\times10^{-4}$ per volume and approximately $0.34\%$ for five independent volumes at $f_{\rm sub}=0.03$ ($1.1\%$ and $2.2\%$ for $f_{\rm sub}=0.1$ and 0.2); reaching 5\% would then require $f_{\rm sub}\simeq0.45$. The $\sim5\%$ figure should therefore not be used as an expectation. The corrected value is closer to the small detection probability of Wang et al.\ \cite{Wang2026}, which is in turn about 20 times smaller than our five-volume estimate, although the two quantities are defined differently and are not directly comparable. This is illustrative, and depends on the assumed sub-halo abundance, spatial distribution, survival and survey geometry, and is distinct from both the Bayes factor and the posterior probability that the candidate is a dark sub-halo.

Wang et al.\ \cite{Wang2026} adopted a different approach, deriving an effective detection volume that incorporates a sub-halo mass function, LOS projection effects, and a requirement that the detectable regions of the two pulsars overlap for a joint detection. Within that framework they obtain a detection probability of $\leq1.7\times10^{-4}$. Analyzing the binary pulsar set (without the solitary pulsars of Ref.~\cite{Donlon_SC_2025}), they find no significant signal under their Galactic-potential assumptions. The potentials they consider include static kinematic potentials (such as \texttt{Galpy} and \texttt{Gala}) derived under equilibrium assumptions that earlier work on pulsar accelerations disfavored \cite{Donlon_SC_2024,Donlon_SC_2025}. Wang et al.\ do find a residual of $2$--$2.6\sigma$ at J1640+2224, one member of the pair identified in Ref.~\cite{Chakrabarti2025}, but discount it on the grounds that the orbital parameters of this source may be degenerate with its parallax. The VLBI astrometric parallax of J1640+2224 \cite{Vigeland2018,Dingeal2023}, which is determined independently of timing, agrees with its timing parallax \cite{EPTA2023DR2}. Moreover, the timing-parallax signature has a semiannual period (182.6 d), close to but distinct from the orbital period ($P_b\simeq175.5$ d), and the two signals beat with a period $[P_b^{-1}-(182.6~\mathrm{d})^{-1}]^{-1}\simeq4.5\times10^3~\mathrm{d}\simeq12$ yr. The present $\sim20$ yr baseline spans about 1.6 beat periods, over which the two signals decorrelate, so the orbital parameters are unlikely to be degenerate with the parallax. Because the two studies use different requirements for sub-halo detection and different population assumptions, they are not directly comparable. In the present work we treat the sub-halo as a \emph{candidate} and condition all microphysical implications on its eventual confirmation.

A separate body of work, largely by Zurek and collaborators, searches for dark matter substructure through the \emph{timing residuals} of pulsar timing arrays (PTAs) rather than through pulsar accelerations \cite{Baghram2011,Dror2019,Ramani2020,LeeMitridate2021,LeeTaylor2021,Cherukupalli2026,Foster2026,Cherukupalli2026Pairs}. Those studies forecast sensitivity to the cosmic abundance of compact and diffuse substructure at masses from $10^{-12}\msun$ up to about $10^{5}\msun$, at least two orders of magnitude below the candidate considered here, and they are expressed as reach in the substructure density of a population rather than as the detectability of one object in existing data. They are therefore not directly comparable with the detection probabilities discussed above. The two approaches measure different observables, in different mass ranges, with different noise. We compare them quantitatively in Sec.~\ref{sec:future-ptas}, where an order-of-magnitude estimate shows that a $\sim10^7\msun$ object at the distance of the candidate leaves only a nanosecond-level residual in a pulsar's timing after the timing model is fitted, so that it is neither detectable nor excluded by such searches (i.e., the work by Zurek and collaborators), while the acceleration observable that we focus on here is essentially unaffected by the gravitational-wave background that limits them.

\section{Reference parameters and the inference framework}
\label{sec:reference}
\subsection{Measured quantities, template assumptions, and mass conventions}
\label{sec:reference-params}
We retain the reported fit normalizations in Table~\ref{tab:reference}, but write $M_{\rm NFW}$ for the mass labeled $M_{\rm vir}$ in the source fit. Operationally it normalizes the adopted profile within $c r_s$; it must not automatically be identified with the present bound mass, a spherical-overdensity mass at a specified epoch, or the peak/infall mass. We use $r_s\equiv R_s$ throughout, and reserve $\sigma_{\rm 1D}$ for a velocity dispersion and $v_c$ for a circular speed.

\begin{table*}[t]
\caption{Reference quantities from Ref.~\cite{Chakrabarti2025} and derived illustrative values. Quoted errors are those of the source fit, for which no covariance matrix or profile uncertainty is available. Derived quantities are not additional measurements.}
\label{tab:reference}
\begin{ruledtabular}
\begin{tabular}{p{0.29\textwidth}p{0.29\textwidth}p{0.34\textwidth}}
Quantity & Value & Interpretation \\
\hline
Compact-object fit mass & $2.45^{+1.07}_{-0.96}\times10^7\msun$ & Effective perturber mass in this model \\
$M_{\rm NFW}$ (source $M_{\rm vir}$) & $6.19^{+1.92}_{-2.03}\times10^7\msun$ & NFW normalization within $c r_s$ \\
$M(<r_s)$ & $0.48^{+0.15}_{-0.16}\times10^7\msun$ & Derived in that NFW model \\
$r_s$, $c$ & $0.1$ kpc, $30$ & Fixed template inputs \\
Galactocentric centroid & $(X,Y,Z)\simeq(7.47,0.38,0.21)$ kpc & NFW-fit location \\
Heliocentric distance $d$ & $\simeq0.78$ kpc & Derived using $R_\odot=8.122$ kpc \\
$\rho_s$, $\rho(r_s)$ & $2.0$, $0.50~\msun\,\mathrm{pc^{-3}}$ & NFW scale and local densities \\
$v_c(r_s)$ & $\simeq14.4~\kms$ & Circular speed, not a measured dispersion \\
$r_t$, $M_t$ (tidal example) & $0.307$ kpc, $1.63\times10^7\msun$ & Spherical point-host approximation \\
Bayes factor & $\sim20$--$40$ & Detection-model comparison, not a DM-model limit \\
\end{tabular}
\end{ruledtabular}
\end{table*}

For the template,
\begin{align}
\rho(r)&=\frac{\rho_s}{x(1+x)^2},\qquad x=r/r_s,\label{eq:nfw}\\
F(x)&\equiv\ln(1+x)-\frac{x}{1+x},\nonumber\\
M(<r)&=4\pi\rho_s r_s^3 F(x),\label{eq:nfw-mass}\\
\rho_s&=\frac{M_{\rm NFW}}{4\pi r_s^3F(c)}
\simeq2.0~\msun\,\mathrm{pc^{-3}},\label{eq:rhos}\\
v_c(r_s)&=\sqrt{\frac{GM(<r_s)}{r_s}}\simeq14.4~\kms.\label{eq:vrms}
\end{align}
Here $\rho_s\simeq1.35\times10^{-22}~\mathrm{g\,cm^{-3}}$, whereas the density at $r_s$ is four times smaller. Neither quantity is a finite NFW central density. The heliocentric distance follows from the three-dimensional centroid and Solar position, not the Galactocentric $X$ difference alone.

\paragraph{A quantitative mass-convention check.}
For a spherical-overdensity (SO) definition,
\begin{align}
M_\Delta(z)&=\frac{4\pi}{3}\Delta\rho_{\rm ref}(z)R_\Delta^3,
&c_\Delta&=R_\Delta/r_s,\nonumber\\
\frac{\rho_s}{\rho_{\rm ref}(z)}&=
\frac{\Delta}{3}\frac{c_\Delta^3}{F(c_\Delta)}.
\label{eq:so-consistency}
\end{align}
Here the reference density is explicitly critical ($\rho_c$) or cosmic mean matter ($\bar\rho_m$); $\rho_c(z)=3H^2(z)/(8\pi G)$. Using $H_0=67.4~\mathrm{km\,s^{-1}\,Mpc^{-1}}$ and $\Omega_m=0.315$ gives $\rho_{c,0}=126.1~\msun\,\mathrm{kpc^{-3}}$. The quoted triplet has $c r_s=3$ kpc and hence
\begin{equation}
\frac{\bar\rho(<3~\mathrm{kpc})}{\rho_{c,0}}=4.34\times10^3,
\qquad
\frac{\bar\rho(<3~\mathrm{kpc})}{\bar\rho_{m,0}}=1.38\times10^4.
\label{eq:observed-overdensity}
\end{equation}
It is therefore not the triplet of a $z=0$ $200c$ or $200m$ halo. This does not invalidate its use as an acceleration template, but fixes what its fitted mass means: $M_{\rm NFW}=M(<3~\mathrm{kpc})$ within that template.

Two different conversions are possible. Preserving the fitted inner $(\rho_s,r_s)$ and formally extending the NFW profile gives $(R_{200c},c_{200c},M_{200c})=(9.47~\mathrm{kpc},94.7,8.96\times10^7\msun)$ and $(R_{200m},c_{200m},M_{200m})=(14.43~\mathrm{kpc},144.3,1.00\times10^8\msun)$. These are mathematical extrapolations, not the sizes or masses of a surviving sub-halo in the Galactic disk. Alternatively, preserving $M_{200}=6.19\times10^7\msun$ and $c_{200}=30$ requires $r_s=0.279$ kpc for $200c$, or $0.410$ kpc for $200m$. That changes the density and acceleration profile and requires a refit. Neither operation is a harmless replacement of the mass label. In flat $\Lambda$CDM, the original triplet formally satisfies the $200c$ relation at $z\simeq3.06$; this density-matching redshift is not a measurement of formation or infall time.

\paragraph{Which radii does an accelerometer measure?}
Let $\mathbf x_h$ be the halo center, $\mathbf x_\odot$ the Solar position, $\mathbf n_i$ the unit vector from the Sun to pulsar $i$, $r_i=|\mathbf x_i-\mathbf x_h|$, and $r_\odot=|\mathbf x_\odot-\mathbf x_h|$. With positive acceleration defined away from the observer, the sub-halo contribution is
\begin{align}
\delta a_i={}&-G\mathbf n_i\cdot\left[
\frac{M(<r_i)(\mathbf x_i-\mathbf x_h)}{r_i^3}
\right.\nonumber\\[-2pt]
&\left.\hspace{3.6em}-
\frac{M(<r_\odot)(\mathbf x_\odot-\mathbf x_h)}{r_\odot^3}
\right].
\label{eq:obs2-forward}
\end{align}
Both endpoint accelerations are required. The Solar contribution is not negligible a priori: for the quoted NFW centroid, its magnitude is $7.36\times10^{-10}~\mathrm{cm\,s^{-2}}$ in the untruncated template and $3.70\times10^{-10}~\mathrm{cm\,s^{-2}}$ in our illustrative tidal model, before projection. The original implementation's sign and Solar-subtraction conventions must be matched explicitly rather than inferred from a displayed acceleration formula.

We computed endpoint radii using the public v3 acceleration catalog accompanying Ref.~\cite{Donlon_SC_2025}, its curated distances, $\mathbf x_\odot=(8.122,0,0)$ kpc, and the fixed NFW centroid in Table~\ref{tab:reference}. Among all 53 catalog sources, the nearest nominal endpoint is J1713+0747 at $r=0.388$ kpc; J1640+2224 is at $0.492$ kpc (Table~\ref{tab:obs2-radii}). No nominal endpoint lies inside 0.1 kpc, or inside the illustrative $r_t=0.307$ kpc. The Sun is at $r_\odot=0.783$ kpc. These radii are deterministic geometry at the nominal distances; Appendix~\ref{app:obs2} describes the coordinate convention, the distance-only variations listed in Table~\ref{tab:obs2-radii}, and the limitations of this calculation.

\begin{table}[t]
\caption{Nearest catalog endpoints at the fixed quoted NFW centroid. Radii are in kpc; the intervals vary only the pulsar distance, as described in Appendix~\ref{app:obs2}.}
\label{tab:obs2-radii}
\begin{ruledtabular}
\begin{tabular}{lcc}
Pulsar & Nominal $r_i$ & Distance-only 16--84\% \\
\hline
J1713+0747 & 0.388 & 0.375--0.399 \\
J1744$-$1134 & 0.425 & 0.421--0.430 \\
J1614$-$2230 & 0.461 & 0.461--0.465 \\
J1640+2224 & 0.492 & 0.351--0.724 \\
J1751$-$2857 & 0.505 & 0.443--0.731 \\
\end{tabular}
\end{ruledtabular}
\end{table}

\begin{figure*}[t]
\centering
\includegraphics[width=0.88\textwidth]{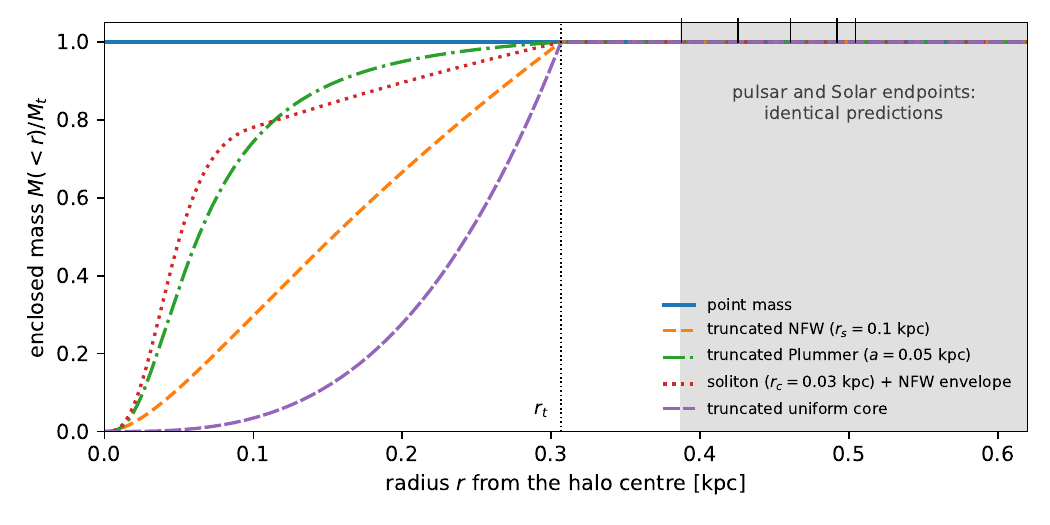}
\caption{An exact degeneracy of the endpoint-acceleration observable at fixed geometry. Five spherical mass models share the same total $M_t$ and are confined within the illustrative $r_t=0.307$ kpc. Their interior enclosed masses differ, but all nominal catalog pulsars and the Sun lie outside this radius, where their acceleration predictions coincide. Pulsar radii within the plotted range are marked above the curves; the shaded range begins at the nearest nominal endpoint. Illustrative parameters: NFW $r_s=0.1$ kpc, Plummer $a=0.05$ kpc, and a soliton with $r_c=0.03$ kpc matched at 0.1 kpc to an NFW envelope with $r_s=0.1$ kpc. These are normalized diagnostic profiles, not fitted density profiles, posterior samples, or verified equilibrium models.}
\label{fig:obs2-identifiability}
\end{figure*}

For any spherical profile confined within a radius smaller than \emph{all} these endpoint radii, the shell theorem reduces Eq.~\eqref{eq:obs2-forward} exactly to the point-mass prediction with mass $M_t$. An arbitrary redistribution of mass inside that boundary leaves every modeled acceleration unchanged. We verified this numerically for a compact mass, a truncated NFW halo, a truncated Plummer core, a soliton plus NFW envelope, and a truncated constant-density core, each normalized to the same $M_t$ (Fig.~\ref{fig:obs2-identifiability}). Thus even perfect nominal-distance accelerations cannot determine their internal scale radii in this fixed-centroid configuration. This is an identifiability result, not evidence that all these profiles are dynamically viable or that a free-centroid fit would remain on the same branch. Section~\ref{sec:future-accel} discusses how endpoints inside the halo break this degeneracy.

The different published compact and NFW normalizations can also be understood without interpreting their ratio as a particle-model preference. In the quoted untruncated NFW template, $M(<0.388~\mathrm{kpc})=1.98\times10^7\msun$ and $M(<0.492~\mathrm{kpc})=2.38\times10^7\msun$; its enclosed mass reaches the compact-fit normalization $2.45\times10^7\msun$ at $r=0.513$ kpc. These scales explain why the two normalizations can differ while producing comparable local forces, but the projected Solar term and joint centroid fit prevent identifying them exactly. Likewise, truncating the \emph{unchanged} NFW density at our illustrative tidal radius leaves only $M_t=1.63\times10^7\msun$. It does not preserve the original best-fit endpoint accelerations. A tidally consistent model must be refitted before its $J$, $D$, or survival times are used as inferred quantities.

\paragraph{From the instantaneous profile to mass history.}
Define $M_t$ as the present bound mass, $M_{\rm infall}$ as the bound mass at first infall, and $M_{\rm peak}=\max_t M_{\rm bound}(t)$ under the same mass convention. Neither of the latter is an algebraic conversion of $M_{\rm NFW}$. Given history variables $\mathcal H$ (initial structure, orbit, infall epoch, host/disk evolution, stripping, and any microscopic evolution), infer them through a forward evolution map $\mathcal E$ and a specified prior:
\begin{align}
p(M_{\rm infall},\mathcal H\mid\mathcal D_a)
\propto{}&p(M_{\rm infall},\mathcal H)\nonumber\\
&\times\mathcal L_a\!\left[\mathcal D_a\mid
\mathcal E(M_{\rm infall},\mathcal H)\right].
\label{eq:obs2-history}
\end{align}
$M_{\rm peak}$ is then derived along each history. With these bound-mass definitions $M_{\rm peak}\geq\max(M_t,M_{\rm infall})$; $M_{\rm infall}\geq M_t$ requires an additional no-net-growth assumption after infall. For illustration, retained fractions $M_t/M_{\rm infall}=1,0.1,0.01$ applied to $M_t=1.63\times10^7\msun$ correspond to $1.63\times10^7,1.63\times10^8,1.63\times10^9\msun$. They are alternative histories, not an inferred interval. A merger-tree mass defined by $M_{200}(z)$ must first be reconciled with this bound-mass convention. Until evolutionary priors and the actual likelihood are supplied, no posterior on $M_{\rm peak}$ or $M_{\rm infall}$ is warranted.

\paragraph{Allowing the scale radius to vary.}
Reference~\cite{Chakrabarti2025} fixed $r_s=0.1$ kpc and $c=30$; these values are of the order expected from cosmological simulations for such objects after tidal truncation \cite{ErkalBelokurov2015,Moline2017}. We repeated its acceleration MCMC with $\log_{10}(r_s/\mathrm{kpc})$ sampled uniformly between $-2$ and $0$, keeping $c=30$ and the same two Galactic potentials (exponential disk, with and without the spiral perturbation). The scale radius is not constrained. The posterior medians, with 16th--84th percentile ranges, are $r_s=61^{+274}_{-44}$ pc for the exponential disk and $74^{+319}_{-55}$ pc with the spiral component, with $M_{\rm NFW}=5.2^{+16.1}_{-2.8}\times10^7\msun$ and $4.5^{+14.6}_{-2.7}\times10^7\msun$, respectively; the two potentials give consistent results. These ranges are comparable to the width of the prior itself (its 16th--84th percentiles are 21 and 480 pc), so they reflect the prior rather than a measurement. The medians are, however, of order a few $10^{-2}$ kpc, comparable to the adopted $0.1$ kpc and well inside the illustrative tidal radius. Two consequences matter below. First, no core size or inner-profile statement follows from the present data, as expected from the exterior geometry above. Second, and more important, the annihilation factor is highly sensitive to the unmeasured $r_s$: at fixed $M_{\rm NFW}$ and for $r_s\lesssim r_t$ it scales roughly as $\rho_s^2r_s^3\propto M_{\rm NFW}^2r_s^{-3}$ (Sec.~\ref{sec:indirect-baseline}).

\subsection{A half-mode scale is not a minimum halo mass}
\label{sec:likelihood}
Define the relative transfer amplitude $T(k)=[P_{\rm model}(k)/P_{\rm CDM}(k)]^{1/2}$ for a specified cosmology and epoch. We distinguish a half-\emph{power} scale, $T^2(k_{\rm hp})=1/2$, from a half-\emph{amplitude} scale, $T(k_{\rm hm}^{(T)})=1/2$. Both are called ``half-mode'' in parts of the literature. With a specified wavenumber-to-mass convention,
\begin{equation}
M(k)=\frac{4\pi}{3}\bar\rho_{m,0}\left(\frac{\pi}{k}\right)^3.
\label{eq:k-mass}
\end{equation}
The density convention and factors of $h$ must accompany every quoted coefficient. Neither definition is a hard lower limit on halo mass, nor is it the mass where a nonlinear sub-halo mass function is exactly halved. Free streaming and wave dynamics suppress formation; they do not generally impose a step function on the surviving population \cite{Schneider2012,Du2017}.

Tidal evolution supplies an additional distinction: a presently small sub-halo can descend from an object formed well above a primordial suppression scale. A small surviving mass therefore does not by itself demonstrate unsuppressed initial power at that mass. The compact-fit mass is not even a halo mass unless that interpretation is separately supported.

A useful population-level starting point is
\begin{align}
\mu_{\rm det}(\theta,\eta)&=\int d\xi\,
\frac{dN_{\rm surv}}{d\xi}(\xi\mid\theta,\eta)
S(\xi\mid\eta),\label{eq:expected-count}\\
P(N\geq1\mid\theta,\eta)&=1-e^{-\mu_{\rm det}(\theta,\eta)},\label{eq:existence-prob}
\end{align}
where $\theta$ denotes microscopic parameters, $\eta$ astrophysical and observational nuisance parameters, $\xi$ the present object properties together with any latent history, and $S$ the search selection function. A marked point-process likelihood also uses the measured properties of each candidate and the search's nondetections. False positives require a background intensity or a reliability mixture inferred jointly, rather than assigning a fixed reliability from the Bayes factor. Counts of pulsar pairs are correlated and are not independent halo detections.

Proximity can make one detection informative, but it is the effective search volume and predicted local abundance, not a sphere of radius $d$ chosen after detection, that set its weight. An object this close may challenge a baseline CDM population as well as a suppressed-power alternative. Wang et al.\ give a concrete example of low detection probabilities under their assumptions \cite{Wang2026}; we do not adopt their value as a universal prior. Even a resolved profile for one object cannot by itself establish a mass-function cutoff.

\subsection{Cosmological assumptions and mixed constituents}
\label{sec:mixed}
Structural observables constrain the combination of primordial fluctuations, transfer physics, and nonlinear evolution. A schematic decomposition is $P_m=\mathcal{T}_{\mathcal R}^2P_{\mathcal R}+P_{\rm iso}$, with cross terms if modes are correlated. Small-scale features, running, non-Gaussian initial conditions, or isocurvature can modify collapse. A nonstandard expansion or production history can alter both perturbation growth and the DM momentum distribution. These changes must be evolved together; enhanced primordial power need not rescue a model whose pressure or subsequent dynamics prevent a bound object. Lensing studies illustrate the use of low-mass structure to constrain primordial power \cite{Gilman2022Power}.

The same caution applies to constituent fractions. A cold majority may form halos despite a warm or wave-like minority. Interacting, unstable, or PBH fractions can have different spatial distributions after formation. We distinguish cosmological fractions from $f_i(r)=\rho_i(r)/\rho_{\rm tot}(r)$ in the present object. The latter need not be constant or equal to the former. A formal half-mode crossing can disappear altogether for sufficiently weak suppression in a mixed model. No universal statement about all particle masses at all subdominant fractions follows from a single candidate.

\emph{Baseline and conditionality.} All numerical values below assume a flat $\Lambda$CDM background with $h=0.674$, $\Omega_m=0.315$, $\Omega_{\rm DM}=0.265$ (so $\Omega_mh^2=0.143$), a standard power-law primordial spectrum without running, and the standard thermal history; warm candidates are specified by the production relations of Sec.~\ref{sec:wdm} (thermal relic and the sterile-neutrino mapping). The matching values are conditional on these choices, and no exclusion based on internal dynamics is derived in this paper. For a warm fraction $f_w$ of the dark matter, perturbations in that component are erased below its free-streaming scale, so at early times the total matter amplitude relative to CDM tends to a plateau of about $1-f_w$; thereafter the cold component grows as $\delta_c\propto a^{p}$ with $p=[\sqrt{1+24(1-f_w)}-1]/4<1$, lowering the plateau further. A half-amplitude crossing therefore exists at fixed epoch only for sufficiently large $f_w$, and its location depends on epoch.

\section{Wave-like (ultralight) dark matter}
\label{sec:fdm}
\subsection{A reproducible suppression-scale estimate}
\label{sec:fdm-hm}
For the canonical non-self-interacting ultralight scalar, Hu, Barkana, and Gruzinov give the illustrative transfer function and half-power scale \cite{Hu2000}
\begin{align}
T_F(k)&\simeq\frac{\cos(x^3)}{1+x^8},\quad
x=1.61m_{22}^{1/18}k/k_{J,\rm eq},\nonumber\\
k_{J,\rm eq}&\simeq9m_{22}^{1/2}~\mathrm{Mpc^{-1}},\nonumber\\
k_{\rm hp}&\simeq4.5m_{22}^{4/9}~\mathrm{Mpc^{-1}}.\label{eq:fdm-k}
\end{align}
Here $m_{22}\equiv m/(10^{-22}~\mathrm{eV})$. Using Eq.~\eqref{eq:k-mass} and the illustrative density $\Omega_mh^2=0.143$ yields
\begin{equation}
M_{\rm hp}\simeq5.7\times10^{10}m_{22}^{-4/3}\msun.
\label{eq:fdm-hm}
\end{equation}
Equating this scale to a comparison mass $M_{\rm match}$ defines
\begin{equation}
m_{\rm match}^{F}=10^{-22}~\mathrm{eV}
\left(\frac{5.7\times10^{10}\msun}{M_{\rm match}}\right)^{3/4}.
\label{eq:fdm-hm-bound}
\end{equation}
For $M_{\rm match}=2.45\times10^7$ and $6.19\times10^7\msun$, this gives about $3.3\times10^{-20}$ and $1.7\times10^{-20}$ eV, respectively. These are algebraic matching values, not exclusion limits or preferred masses.

The scale of Eqs.~\eqref{eq:fdm-k}--\eqref{eq:fdm-hm} is the half-power scale of the Hu et al.\ fit and should not be confused with the soliton scale definition of Marsh and Pop \cite{MarshPop2015}; an HMF-calibrated suppression scale must be introduced with a different explicit definition. For a scaling $M_{\rm hp}=C\,m_{22}^{-4/3}$, with $C=5.7\times10^{10}\msun$ in Eq.~\eqref{eq:fdm-hm}, $d\ln m_{\rm match}^{F}=(3/4)(d\ln C-d\ln M_{\rm match})$; this derivative propagates a change of convention or of comparison mass, but it is a systematic choice rather than a statistical uncertainty and supplies no confidence interval.

Canonical $m\sim10^{-22}$ eV scalar DM strongly suppresses the initial scales relevant to low-mass halos. Comparing a confirmed object with that prediction nevertheless requires its progenitor distribution and the selection likelihood of Sec.~\ref{sec:likelihood}. Published benchmarks include $m>2\times10^{-20}$ eV from a canonical-scalar Lyman-$\alpha$ analysis and $m>2.9\times10^{-21}$ eV from the cited satellite analysis, both at 95\% confidence in their respective models \cite{RogersPeiris2021,Nadler2021}. These provide context, not a significance calibration for this candidate.

Solving $T_F^2(k_{\rm hp})=1/2$ exactly with Eq.~\eqref{eq:fdm-k} gives $k_{\rm hp}=4.58$, 12.76, and $35.5~\mathrm{Mpc^{-1}}$ at $m_{22}=1$, 10, and 100, about 2\% above the fitted scaling $4.5\,m_{22}^{4/9}$, i.e.\ $M_{\rm hp}=5.35\times10^{10}$, $2.48\times10^{9}$, and $1.15\times10^{8}\msun$ against the fitted $5.66\times10^{10}$, $2.63\times10^{9}$, and $1.22\times10^{8}\msun$. Halving the amplitude $T_F$ instead gives $k=5.08~\mathrm{Mpc^{-1}}$ and $M=3.9\times10^{10}\msun$ at $m_{22}=1$, which would lower the matching masses to about $2.5\times10^{-20}$ and $1.3\times10^{-20}$ eV. Throughout, the mass filter is the sharp-$k$ sphere of radius $\pi/k$ of Eq.~\eqref{eq:k-mass}.

\subsection{Soliton size and density}
\label{sec:fdm-soliton}
An often-used $z=0$ core--halo scaling is \cite{SchivePRL2014}
\begin{equation}
r_c\simeq1.6~\mathrm{kpc}\,m_{22}^{-1}
\left(\frac{M_{\rm halo}}{10^9\msun}\right)^{-1/3}.
\label{eq:soliton}
\end{equation}
Inserting $M_{\rm halo}=6.2\times10^7\msun$ gives $r_c\simeq4.0~\mathrm{kpc}/m_{22}$. If a future measurement established $r_c<0.1$ kpc and this scaling were applicable, the corresponding condition would be $m\gtrsim4\times10^{-21}$ eV. The fixed NFW scale radius is not such a measurement, and it is not the same profile parameter as a soliton half-density radius. Core--halo scatter, formation history, stripping of the envelope, and mixed components further weaken a direct translation.

For a resolved soliton, $\rho_{\rm sol}(r)=\rho_c[1+0.091(r/r_c)^2]^{-8}$ and $\rho_c\propto m^{-2}r_c^{-4}$ give a complementary test \cite{Schive2014,MarshPop2015}. Measurements of $\rho_c$ and $r_c$ determine one mass under this profile assumption; an independent envelope/halo measurement is needed to test the core--halo relation. They do not alone give two independent measurements of $m$. Ultralight self-interactions or multicomponent wave dynamics require an altered soliton model.

\subsection{Granularity and coherent pulsar-timing signals}
\label{sec:fdm-granularity}
Using the reduced de Broglie wavelength,
\begin{equation}
\lambda_{\rm dB}=\frac{\hbar}{mv}\simeq0.19~\mathrm{kpc}\,
m_{22}^{-1}\left(\frac{v}{100~\kms}\right)^{-1},
\label{eq:lambda-dB}
\end{equation}
the rough ambient granule acceleration is
\begin{equation}
\delta a\sim G\rho_{\rm loc}\lambda_{\rm dB},\qquad
\tau_{\rm dB}\sim\lambda_{\rm dB}/v.
\label{eq:granule-accel}
\end{equation}
For $m=4\times10^{-21}$ eV, $v=200~\kms$, and $\rho_{\rm loc}=0.3~\mathrm{GeV\,cm^{-3}}$, this gives $\lambda_{\rm dB}\simeq2.4$ pc, $\delta a\simeq2.6\times10^{-13}~\mathrm{cm\,s^{-2}}$, and $\tau_{\rm dB}\simeq1.2\times10^4$ yr. The density is that of the ambient halo, not the sub-halo center \cite{HuiOstrikerTremaineWitten2017,Read2014,Donlon_SC_2025}. This illustrative fluctuation is approximately $2300$ times below the $6\times10^{-10}~\mathrm{cm\,s^{-2}}$ selection threshold and decorrelates over scales much smaller than the $0.3$--$0.5$ kpc pulsar separations. This length is about 40 times smaller than $r_s$. Acceleration fluctuations projected along actual sightlines require their correlation function rather than a single-cell estimate.

A distinct signal is the coherent oscillating potential, with frequency $f\simeq mc^2/(\pi\hbar)=48~\mathrm{nHz}\,m_{22}$ \cite{KhmelnitskyRubakov2014}. At $m=4\times10^{-21}$--$10^{-20}$ eV its period is approximately 6--2.4 days, unrelated to the much longer granule crossing time. Existing scalar searches have assumption-dependent density constraints \cite{Smarra2023EPTA,PPTA2026}; the recent PPTA/EPTA analysis does not exclude all-DM scalar abundance under its adopted treatment. Stochastic metric fluctuations probe a different frequency-to-mass relation and have been proposed as a sensitive local-density probe around $10^{-18}$--$10^{-16}$ eV \cite{KimMitridate2024}. Neither result justifies a universal claim about every allowed mass and mixture. These signals probe the ambient field; the distinct PTA searches for transiting substructure are discussed in Sec.~\ref{sec:future-ptas}.

\section{Thermal and nonthermal warm dark matter}
\label{sec:wdm}
\subsection{Transfer-function convention and scale matching}
A reproducible thermal-relic example uses \cite{Viel2005,Schneider2012}
\begin{align}
T_W(k)&=[1+(\alpha k)^{2\nu}]^{-5/\nu},\quad\nu=1.12,\label{eq:wdm-transfer}\\
\alpha&=0.049~h^{-1}\mathrm{Mpc}
\left(\frac{m_{\rm WDM}}{1~\mathrm{keV}}\right)^{-1.11}\nonumber\\
&\quad\times\left(\frac{\Omega_{\rm WDM}}{0.25}\right)^{0.11}
\left(\frac{h}{0.7}\right)^{1.22}.\nonumber
\end{align}
For $T=1/2$, $k_{\rm hm}^{(T)}=\alpha^{-1}(2^{\nu/5}-1)^{1/(2\nu)}$. Taking $h=0.674$, $\Omega_{\rm WDM}=0.265$, and $\Omega_mh^2=0.143$, Eq.~\eqref{eq:k-mass} gives
\begin{equation}
M_{\rm hm}^{(T)}\simeq1.9\times10^{10}\msun
\left(\frac{m_{\rm WDM}}{1~\mathrm{keV}}\right)^{-3.33}.
\label{eq:wdm-hm}
\end{equation}
Equating this scale to $2.45\times10^7$ or $6.19\times10^7\msun$ gives matching masses of approximately $7.4$ or $5.6$ keV. Under the \emph{half-power} definition used in Sec.~\ref{sec:fdm-hm}, the same fit instead gives $M_{\rm hp}\simeq5.1\times10^{10}\msun(m_{\rm WDM}/\mathrm{keV})^{-3.33}$ and matching masses of approximately $9.9$ or $7.5$ keV. The difference reflects the convention alone. Differently normalized expressions in the literature, such as $3\times10^8\msun(m_{\rm WDM}/3.3~\mathrm{keV})^{-3.33}$, correspond to a $1$ keV coefficient of $1.6\times10^{10}\msun$; their assumptions must be matched before they are used alongside Eq.~\eqref{eq:wdm-hm}.

Historical benchmarks include $m_{\rm WDM}>3.3$ keV at $2\sigma$ from Viel et al., and $>5.3$ keV at $2\sigma$ in the combined analysis of Ir\v{s}i\v{c} et al.; the latter relaxes to 3.5 keV with more flexible temperature evolution \cite{Lyman2013,Irsic2017WDM}. The cited satellite analysis finds $m_{\rm WDM}>6.5$ keV at 95\% confidence \cite{Nadler2021}. These limits rest on full likelihoods and astrophysical modeling. Our scale matching is not another confidence-level limit and cannot by itself exclude a portion of their allowed interval. Correlated mass and profile errors and discrete calibration choices should be propagated through a calibrated selection function and population model (Sec.~\ref{sec:likelihood}).

With $h=0.674$, $\Omega_{\rm WDM}=0.265$, and $\Omega_mh^2=0.143$, Eq.~\eqref{eq:wdm-transfer} gives $\alpha=0.0471\,h^{-1}\mathrm{Mpc}$ at 1 keV. The half-amplitude condition $T=1/2$ then gives $k=6.45~\mathrm{Mpc^{-1}}$ and $M=1.92\times10^{10}\msun$, and the half-power condition $T^2=1/2$ gives $k=4.65~\mathrm{Mpc^{-1}}$ and $5.1\times10^{10}\msun$; the mass exponent is $3\times1.11=3.33$. These coefficients yield the matching masses quoted above and those for $10^6\msun$ below. The fit is a $z=0$ linear transfer function, used here only to define a scale.

\subsection{Production mechanisms and sterile-neutrino translations}
For a nonresonantly produced sterile neutrino, a commonly used thermal-equivalent mapping is \cite{ColombiDodelsonWidrow1996,Viel2005}
\begin{equation}
m_s\simeq4.43~\mathrm{keV}
\left(\frac{m_{\rm WDM}}{1~\mathrm{keV}}\right)^{4/3}
\left(\frac{\Omega_{\rm DM}h^2}{0.1225}\right)^{-1/3}.
\label{eq:sterile-map}
\end{equation}
An alternative production calculation gives an approximate $3.90~\mathrm{keV}(m_{\rm WDM}/\mathrm{keV})^{1.294}$ scaling at the reference abundance \cite{Abazajian2006}. At $m_{\rm WDM}=6$ keV these yield roughly 48 and 40 keV, illustrating production dependence. A resonantly produced or decay-produced distribution can be colder, nonthermal, or multipeaked; matching one suppression scale is then insufficient. The same point applies to freeze-in relics, gravitinos, and other warm candidates: the momentum distribution and abundance, not mass alone, control their transfer function. X-ray limits on sterile neutrinos constrain the mixing angle and must be imposed jointly with the production relation, which ties that angle to the abundance.

If future data were sensitive to objects corresponding to a \emph{formation} scale of $10^6\msun$, half-amplitude matching in Eq.~\eqref{eq:wdm-hm} would reach about 19 keV (about 26 keV under half-power matching). This is a scaling statement; it does not establish that such pulsar sensitivity is achievable. A $10^6\msun$ stripped remnant need not have that formation scale.

\section{Interactions that alter formation and halo evolution}
\label{sec:sidm}
\subsection{Dark acoustic oscillations and Standard Model scattering}
DM coupled to dark radiation, a light mediator, or Standard Model particles can inherit damping and acoustic features. ETHOS provides an effective description linking such microphysics to transfer functions and nonlinear structure \cite{CyrRacine2016,Vogelsberger2016,Buckley2014}. Elastic self-scattering by itself does not require a primordial cutoff: early kinetic coupling and late halo scattering are distinct physical processes, even when one mediator produces both.

An oscillatory transfer function is not specified by its first half-mode crossing. Oscillation amplitude, damping envelope, later peaks, and the thermal history can affect abundance, concentrations, and collapse epochs at fixed nominal cutoff. The WDM template is therefore a comparison tool, not a universal ETHOS HMF. Likewise, scattering with baryons, including DM--proton scattering, can suppress small-scale structure and connects halo demographics to a directly particle-physics interaction \cite{Nadler2019Baryon}. This is distinct from present-day internal SIDM heat transport.

\subsection{Internal scattering: a timescale diagnostic}
\label{sec:sidm-core}
For an approximately isotropic, velocity-independent collision cross section, a local scattering rate is
\begin{equation}
\Gamma(r)=\rho(r)\left\langle\frac{\sigma(v_{\rm rel})}{m_\chi}v_{\rm rel}\right\rangle.
\label{eq:scatter-rate}
\end{equation}
For a Maxwellian with one-dimensional dispersion $\sigma_{\rm 1D}$, $\langle v_{\rm rel}\rangle=4\sigma_{\rm 1D}/\sqrt\pi$. A mean collision time is not automatically a calibrated core-growth or heat-conduction time. For anisotropic scattering one must also specify the transfer/viscosity weighting used by the chosen transport model.

Using $\rho_s$ and $v_c(r_s)$ only as dimensional scale quantities gives
\begin{equation}
t_s\equiv[\rho_s(\sigmam)v_c(r_s)]^{-1}
\simeq0.16~\mathrm{Gyr}\,
\frac{1~\mathrm{cm^2\,g^{-1}}}{\sigmam}.
\label{eq:tcore}
\end{equation}
At the physical radius $r_s$, using instead $\rho(r_s)=\rho_s/4$ and, purely illustratively, $\sigma_{\rm 1D}=15~\kms$, the Maxwellian collision time is approximately $0.28~\mathrm{Gyr}(1~\mathrm{cm^2\,g^{-1}})/(\sigmam)$ (0.42 and 0.21 Gyr for $\sigma_{\rm 1D}=10$ and $20~\kms$, respectively). Neither timescale is a calibrated exclusion. Requiring $t_s$ to exceed an assumed 5--10 Gyr yields a dimensional scale of order $10^{-2}~\mathrm{cm^2\,g^{-1}}$, but neither the halo age nor a lack of profile evolution has been measured. Since every nominal pulsar endpoint lies outside the tidal radius (Sec.~\ref{sec:reference}), the present acceleration data cannot supply the core-size or profile-change measurement that a no-evolution argument would require.

\subsection{Core expansion, collapse, and dense survivors}
\label{sec:sidm-collapse}
Elastic SIDM can first conduct heat into a cold central region, creating an expanded core, and subsequently enter gravothermal collapse with a shrinking, denser center. A dense object can therefore occupy either a weakly evolved branch or a strongly evolved collapsed branch. Stripping of the outer heat reservoir can accelerate collapse; a long isolated-halo collapse time is not sufficient to discard the latter possibility \cite{Nishikawa2020,Zeng2025}. Large cross sections may also enter transport regimes in which a simple inverse-cross-section timescale ceases to apply.

The inference from one compact survivor is consequently nonmonotonic in cross section. A measured profile could disfavor an expanded-core model while remaining compatible with a collapsed SIDM population. Tidal survivability and central density must be predicted together, with initial concentration, orbit, host interactions, and assembly history marginalized. A measured core radius is not in general a direct algebraic measurement of $\sigmam$, even when $r_c=r_s$.

Velocity-dependent models can separate internal dwarf-scale scattering from much faster encounters with the host. Benchmark dwarf and cluster cross sections \cite{KaplinghatTulinYu2016,TulinYu2018,Sagunski2021} cannot be compared as though they were the same observable at the same velocity. The proposed object's $10$--$20~\kms$ internal scale is useful but not an otherwise unconstrained regime: dwarf and ultrafaint systems already inform low-velocity models. A genuinely dark object would supply different baryonic systematics rather than exclusive access to that velocity range.

\subsection{Tidal truncation and the host environment}
\label{sec:sidm-tidal}
For a circular orbit in a spherical host, a useful Jacobi-radius approximation is
\begin{equation}
r_t=R\left[\frac{M(<r_t)}{A(R)M_{\rm MW}(<R)}\right]^{1/3},
\quad A=3-\frac{d\ln M_{\rm MW}}{d\ln R}.
\label{eq:tidal-radius}
\end{equation}
The factor is 3 for a point-mass host and 2 for a flat circular-speed curve. With $R=8.1$ kpc, $M_{\rm MW}(<R)=10^{11}\msun$, and $A=3$, solving Eqs.~\eqref{eq:nfw-mass} and \eqref{eq:tidal-radius} gives
\begin{equation}
r_t\simeq0.307~\mathrm{kpc},\qquad M_t\simeq1.63\times10^7\msun.
\label{eq:tidal-numbers}
\end{equation}
As demonstrated in Sec.~\ref{sec:reference}, truncating this fixed NFW normalization changes the accelerations at the measured endpoints; Eq.~\eqref{eq:tidal-numbers} is not a refitted tidal solution. The mean host density is $0.045~\msun\,\mathrm{pc^{-3}}$; at the adopted tidal boundary the mean sub-halo density is $A$ times that value. For an isothermal host the mean enclosed density is $3\rho_{\rm host}(R)$ and $A=2$, so $\bar\rho_{\rm sub}(<r_t)\simeq6\rho_{\rm host}(R)$, a condition independent of $r_s$. Comparing the NFW scale density to the host mean gives a ratio near 45, but this compares a template parameter to the host density; it is not a measured central-density safety margin or a proof of survival through repeated shocks. The dependence of $r_t$ on $r_s$ is also conditional on which mass or density parameters are held fixed. As a check of the host assumption, a flat circular-speed curve ($A=2$, $V_c=233.4~\kms$) at the Galactocentric radius of the centroid, $R\simeq7.48$ kpc ($M_{\rm MW}(<R)=9.5\times10^{10}\msun$), gives $r_t\simeq0.34$ kpc and $M_t\simeq1.78\times10^7\msun$, while a point-mass host ($A=3$) at that radius gives $0.28$ kpc and $1.51\times10^7\msun$. Over these hosts $M_t$ therefore varies by less than about 10\% around the reference value, and the $A=2$ radius remains smaller than the nearest nominal endpoint (0.388 kpc).

The candidate lies near the disk and at Galactocentric radius approximately 7.5 kpc; the $R=8.1$ kpc example above is retained as a clearly labeled Solar-circle approximation. A Galactic disk is not spherical, and eccentric orbits require a tidal tensor, impulsive versus adiabatic heating, and mass-loss history. Core expansion can enhance disruption, whereas stripping can accelerate SIDM collapse and produce denser survivors.

Host--sub-halo collisions additionally produce evaporation, heat exchange, and drag. Schematically,
\begin{equation}
\Gamma_{\rm evap}\sim\rho_{\rm host}
\left\langle\frac{\sigma}{m_\chi}v_{\rm rel}
P_{\rm ej}(v_{\rm rel},v_{\rm esc})\right\rangle,
\label{eq:host-evap}
\end{equation}
where the ejection probability includes angular and escape-speed dependence. Relevant velocities include host dispersion and orbital motion, often hundreds of $\kms$, separately from the internal velocities in Eq.~\eqref{eq:scatter-rate} \cite{Zeng2025}. A small systemic speed relative to the disk is not necessarily a small speed relative to halo DM.

\subsection{Inelastic and dissipative self-interactions}
\label{sec:inelastic}
Exothermic transitions can inject enough kinetic energy to eject particles even when elastic encounters at the same cross section would mainly redistribute heat. Endothermic transitions have thresholds and remove kinetic energy; additional radiative or dissipative channels can accelerate contraction if the energy escapes. The excited-state fraction, splitting, replenishment and decay rates, and whether emitted dark radiation is trapped all matter \cite{Vogelsberger2019Inelastic,Essig2019}. A dissipative subcomponent might develop morphology unlike a spherical halo; one should fit that possibility only for an explicit cooling model.

\section{Invisible decays and recoil-driven evolution}
\label{sec:invisible}
\subsection{The lifetime--kick--fraction parameter space}
A structurally important decay need not produce photons. Consider $X\to Y+\phi$, with a stable massive daughter $Y$ and invisible radiation $\phi$. In the parent rest frame, for $m_\phi=0$, defining $\epsilon=(m_X-m_Y)/m_X$ gives
\begin{equation}
\frac{v_k}{c}=\frac{m_X^2-m_Y^2}{m_X^2+m_Y^2}
\simeq\epsilon\quad(\epsilon\ll1).
\label{eq:decay-kick}
\end{equation}
The relevant parameters include $(\tau_X,v_k,f_X)$, where the reference epoch for $f_X$ must be stated. Kicks below escape speed can heat and expand a halo; larger kicks remove daughters and enhance vulnerability to the host tide. The decay is per particle, whereas scattering rates depend on the partner density. The fraction decayed in an interval $\Delta t$ is $1-e^{-\Delta t/\tau_X}$ for the parents present at its start, but halo response depends on decays before and after infall and whether mass loss is adiabatic.

For the truncated NFW reference, an isolated escape-to-infinity estimate is
\begin{equation}
v_{\rm esc}^2(r)=8\pi G\rho_s r_s^2
\left[\frac{\ln(1+x)}{x}-\frac{1}{1+x_t}\right],
\label{eq:escape}
\end{equation}
with $x_t=r_t/r_s$. It gives about 31 $\kms$ at $r_s$ and 40 $\kms$ at the center. Escape through the Galactic tidal boundary can require less energy. Kicks of tens of $\kms$ therefore warrant investigation; the relevant comparison is with these escape speeds rather than with the circular speed $v_c(r_s)\simeq14~\kms$.

Cosmological simulations and satellite populations already constrain recoil-decay models \cite{Wang2014Decay,Mau2022}. As a benchmark, Mau et al.\ report 95\% limits of $\tau_X>18$ and 29 Gyr at kicks of 20 and 40 $\kms$ in their model and analysis \cite{Mau2022}. These are not limits from the present candidate and do not transfer unchanged to a subdominant unstable species. They probe a very different lifetime range from electromagnetic decay searches at $\sim10^{28}$ s.

\subsection{Possible direct-detection complementarity}
Invisible daughters can still scatter weakly in a terrestrial detector. In that case structural evolution and direct detection test different combinations of lifetime, abundance, splitting, and portal coupling. Nuclear scattering is not implied by invisibility, and its coupling need not set the decay width. Section~\ref{sec:lz} separates a mild decay kick from a genuinely boosted incident population.

\section{Primordial black holes and massive constituents}
\label{sec:pbh}
\subsection{A many-PBH halo: formation and discreteness}
A cold ensemble of many PBHs can form extended halos. A gravitational detection of such a halo does not identify its constituent as an elementary particle. Unlike microscopic CDM, a macroscopic population contributes number-count fluctuations. For uncorrelated, monochromatic PBHs, neglecting baryon weighting for this DM-normalized expression,
\begin{equation}
P_{\rm shot,DM}(k,z)=\mathcal{T}_{\rm iso}^2(k,z)
\frac{f_{\rm PBH}M_{\rm PBH}}{\bar\rho_{\rm DM,0}}.
\label{eq:pbh-shot}
\end{equation}
Here $f_{\rm PBH}$ is the cosmological DM fraction and $\mathcal{T}_{\rm iso}$ evolves the initial Poisson contribution. The factor $f_{\rm PBH}$ follows from weighting the PBH density contrast by its fraction: $f_{\rm PBH}^2/\bar n_{\rm PBH}=f_{\rm PBH}M_{\rm PBH}/\bar\rho_{\rm DM,0}$. This extra power can enhance early low-mass structure rather than impose a WDM-like suppression \cite{Afshordi2003,Colazo2025}. Initial clustering and correlations alter the Poisson assumption.

For a number-weighted mass distribution, the relevant mass moment is $M_{\rm eff}=\langle M^2\rangle_n/\langle M\rangle_n$, not an arbitrary characteristic PBH mass. Shot noise, subsequent accretion, and nonlinear clustering must be incorporated in the same cosmology as other DM components. A measured abundance or concentration distribution could then constrain $(f_{\rm PBH},M_{\rm eff})$, but one uncalibrated nearby candidate does not give a generic PBH fraction limit.

\subsection{Relaxation, mass segregation, and evaporation}
Massive constituents also change collisional gravitational evolution. For a roughly single-mass, nonrelativistic population a local estimate is
\begin{equation}
t_{\rm rel}\sim\frac{0.34\,\sigma_{\rm 1D}^3}
{G^2\rho_{\rm PBH}M_{\rm PBH}\ln\Lambda}.
\label{eq:pbh-relax}
\end{equation}
The numerical factor is distribution- and convention-dependent. In a mixed system, gravitational diffusion involves $\int dM\,n(M)M^2$, species-dependent velocity distributions, and the smooth component. Encounters can redistribute energy, segregate massive PBHs, produce a dense core or binaries, and eventually eject mass; hierarchical assembly and the external tide also matter \cite{Afshordi2003}.

Requiring an unchanged collisionless profile is stronger than requiring that some bound object survive. Relaxation does not imply instantaneous evaporation, and relaxation, core-collapse, and evaporation times are distinct. The age, radius, velocity dispersion, mass spectrum, binary heating, and boundary conditions must be supplied before converting Eq.~\eqref{eq:pbh-relax} into a constituent-mass constraint. Constraints based on heating a stellar population cannot be imported into a genuinely starless object without identifying suitable tracers.

\subsection{Single compact objects and seeded envelopes}
The compact-object acceleration fit also permits a logically different possibility: a single black hole or a compact cluster, rather than a conventional cosmological low-mass halo. Its mass is not a test of the primordial halo cutoff at the same mass. Establishing a black hole would still not establish primordial origin. A PBH surrounded by an accreted particle-DM envelope is a third case, with seed-driven formation and a potentially non-NFW profile. Any annihilation enhancement in that envelope requires its own tidal and particle-physics calculation.

\subsection{Hawking emission and superheavy particle candidates}
Low-mass PBHs with appreciable Hawking emission can supply photons and electrons/positrons. For monochromatic PBHs tracing a halo with constant fraction, the prompt flux is
\begin{equation}
\frac{d\Phi_\gamma}{dE}=
\frac{f_{\rm PBH}D}{4\pi M_{\rm PBH}}
\frac{d^2N_\gamma}{dE\,dt}.
\label{eq:pbh-hawking}
\end{equation}
Here $D$ is the decay factor of Eq.~\eqref{eq:Dfactor} and $d^2N_\gamma/dE\,dt$ is the Hawking emission rate of a single black hole. An extended mass function requires integration over the mass-dependent Hawking spectrum and spatial distribution. This is a density-linear channel, unlike annihilation, and can connect to the transport calculation in Sec.~\ref{sec:indirect-multiwavelength}. Dwarf-galaxy radio studies provide examples of such searches \cite{Liu2024PBH}; no competitive limit for the present target is assumed.

For superheavy \emph{elementary} particles, the existence of a halo of this mass supplies no generic new upper particle-mass bound. Its implications enter through production, momentum distribution, interactions, stability, or detectable emission. The discreteness effects above become relevant to macroscopic constituents rather than merely to particles heavier than conventional WIMPs. Likewise, a dark sub-halo is compatible with many cold particle candidates whose characteristic formation cutoff is far below the probed scale.

\section{Indirect detection: emission, profiles, and transport}
\label{sec:indirect}
The dedicated analysis of Zhu, Liang, and Huang \cite{Zhu2025} supplies an existing observational baseline for prompt annihilation. Its limits, spatial templates, and background tests should be credited and reused where applicable. Independent gravitational information can improve source modeling, but it does not eliminate profile, distance, or mass uncertainties. A cored halo, collapsed cusp, compact object, or seeded envelope can predict very different emission at similar accelerations. We condition each emission calculation on an explicit constituent and density model and propagate its shared uncertainties with the structural analysis.

\subsection{Consistent \texorpdfstring{$J$ and $D$}{J and D} factors}
\label{sec:indirect-jd}
The observational definitions over a specified aperture are
\begin{align}
J(\Delta\Omega)&=\int_{\Delta\Omega}d\Omega\int ds\,\rho^2(r),\label{eq:Jfactor}\\
D(\Delta\Omega)&=\int_{\Delta\Omega}d\Omega\int ds\,\rho(r),\label{eq:Dfactor}
\end{align}
where $s$ is distance from the observer and $r^2=s^2+d^2-2sd\cos\psi$. Written as volume integrals over the whole source, they carry a weight $1/s^2$ that varies across the source. For a source much smaller than its distance the far-field expressions are
\begin{align}
J_{\rm far}&=\frac{4\pi\rho_s^2r_s^3}{3d^2}
\left[1-(1+x_t)^{-3}\right],\label{eq:Jfar}\\
D_{\rm far}&=M_t/d^2.\label{eq:Dfar}
\end{align}
Here $x_t=r_t/r_s$ is the cutoff of the tidally truncated model. An untruncated $c r_s=3$ kpc halo would include the observer at $d=0.78$ kpc and cannot use a distant-source whole-halo approximation.

For any spherical profile with $r_t<d$, an exact whole-source expression useful for checking the line-of-sight integration is
\begin{equation}
I_n=\frac{2\pi}{d}\int_0^{r_t}dr\,r\rho^n(r)
\ln\!\left(\frac{d+r}{d-r}\right),\quad (I_1,I_2)=(D,J).
\label{eq:exact-spherical}
\end{equation}
Using the fiducial normalization, $d=0.78$ kpc, and Eq.~\eqref{eq:tidal-numbers}, numerical integration gives
\begin{align}
J_{\rm all}&\simeq1.21\times10^{23}~\mathrm{GeV^2\,cm^{-5}},\nonumber\\
D_{\rm all}&\simeq3.19\times10^{21}~\mathrm{GeV\,cm^{-2}}.\label{eq:JD-values}
\end{align}
The corresponding far-field values are $1.20\times10^{23}$ and $3.14\times10^{21}$ in the same units. Truncation changes $J$ little for a fixed inner NFW profile but substantially lowers $D$ relative to the untruncated far-field value $M_{\rm NFW}/d^2\simeq1.2\times10^{22}~\mathrm{GeV\,cm^{-2}}$. The whole truncated source subtends a radius of approximately $23^\circ$, whereas $r_s/d$ corresponds to about $7^\circ$. A finite analysis region collects only part of these integrals.

The illustrative $J$ is conditional on retaining the inner NFW normalization despite the change in endpoint accelerations established in Sec.~\ref{sec:reference}; it must be recomputed after a tidal refit. It remains very large compared with commonly quoted dwarf values \cite{McDaniel2024}, but comparisons must use matched apertures and foreground sensitivity; integrated $J$ alone is not a sensitivity ranking. The compact-object alternative does not supply the same particle-annihilation $J$ factor at all. A mixture of profile hypotheses need not have a lognormal $J$ posterior.

For comparison, Zhu et al.\ report $J_{\rm ROI}=1.14\times10^{23}~\mathrm{GeV^2\,cm^{-5}}$ at $d=0.78$ kpc, integrated over their $20^\circ\times20^\circ$ region \cite{Zhu2025}, rather than over the whole source as in Eq.~\eqref{eq:JD-values}. They also report $0.79\times10^{23}$ and $1.82\times10^{23}$ for the distance variants. Their nominal ROI value is about $94\%$ of our illustrative whole-source value, but this comparison does not isolate an aperture correction: their template and our explicit tidal truncation must first be matched. Their distance variants alter both brightness and angular morphology; treating that analysis as a fixed-template, normalization-only uncertainty would misrepresent it. As a numerical check, integrating our tidally truncated template over a $20^\circ\times20^\circ$ square centered on the halo gives $1.15\times10^{23}~\mathrm{GeV^2\,cm^{-5}}$ (a $10^\circ$-radius circle gives $1.14\times10^{23}$), within about $1\%$ of their nominal value; because the cusp dominates $J$, the dependence on aperture is weak, but this does not replace matching their templates.

\subsection{Existing Fermi-LAT analysis and its scope}
\label{sec:indirect-baseline}
Reference~\cite{Zhu2025} analyzes Pass~8 SOURCE-class events from 2008 August 4 through 2025 November 20, in the 0.5--500 GeV range. Its $20^\circ\times20^\circ$ ROI is centered at $(\mathrm{R.A.},\mathrm{Decl.})=(267.78^\circ,4.70^\circ)$, or $(l,b)=(30.23^\circ,15.56^\circ)$. The binned likelihood uses $0.1^\circ$ pixels, 30 logarithmic energy bins, the P8R3\_SOURCE\_V3 response, 4FGL-DR4 sources within $17^\circ$, and the standard Galactic and isotropic diffuse templates. Catalog-source spectral parameters within $8^\circ$, diffuse normalizations, and selected bright-source normalizations outside that radius are free.

With $r_s=0.1$ kpc and $c=30$ fixed, Zhu et al.\ construct separate spatial templates at $d=0.61$, $0.78$, and $0.95$ kpc. Within each template, they profile a lognormal $J$-normalization nuisance with $\sigma_{\log_{10}J}=0.29$ dex, propagated from the mass uncertainty using $J\propto M_{\rm NFW}^2$. The near/far results form a distance-variation band; this is not a simultaneous confidence region over all structural uncertainties. The treatment accounts for changing angular size with distance, but does not explore free $r_s$, concentration, cores, collapsed cusps, tides, or the full joint dynamical posterior. This omission matters: letting $r_s$ vary in the acceleration fit leaves it unconstrained, $r_s=61^{+274}_{-44}$ pc (free-$r_s$ fit of Sec.~\ref{sec:reference-params}), and since $J_{\rm far}\propto\rho_s^2r_s^3\propto M_{\rm NFW}^2r_s^{-3}$ at fixed $M_{\rm NFW}$ for $r_s\lesssim r_t$, the unexplored $r_s$ dependence can exceed the $0.29$ dex mass nuisance.

For self-conjugate annihilation into $b\bar b$ and $\tau^+\tau^-$, the study uses PPPC4DMID photon yields with electroweak corrections and scans $m_\chi=10$ GeV--10 TeV. At each fixed mass it profiles the LAT likelihood times the $J$ nuisance term and obtains a one-sided 95\% upper limit from
\begin{equation}
-2\ln\!\left[
\frac{\mathcal L(m_\chi,\mu_{95},\widehat{\boldsymbol\eta}_{\mu_{95}})}
{\mathcal L(m_\chi,\widehat\mu,\widehat{\boldsymbol\eta})}
\right]=2.71,\qquad \mu=\langle\sigma v\rangle.
\label{eq:zhu-upper-limit}
\end{equation}
Here $\boldsymbol\eta$ denotes nuisance parameters and the denominator is maximized at the \emph{same mass}, allowing $\widehat\mu>0$. The baseline retains the unexplained residual rather than masking it or subtracting an added source. Their Fig.~5 reports limits stronger than the compared stacked-dwarf and blind sub-halo search results and comparable to nominal Ursa Major~III limits, conditional on the adopted NFW model. Their large inferred $J$, rather than an exceptionally clean foreground, drives this comparison. Those curves are results of Ref.~\cite{Zhu2025}; we have not performed a new LAT fit or established a new exclusion here.

Zhu et al.\ also present a substantial residual analysis (their Figs.~1--4). The residual is concentrated below 2 GeV, includes localized and elongated features, and does not closely follow the broad, approximately symmetric NFW morphology. The strongest localized feature lies about $1.5^\circ$ from the dynamical centroid and has a point-source TS of roughly 50. The maximum NFW-template likelihood improvements are separately $\mathrm{TS}_{\rm DM}\simeq60$ for $b\bar b$ and $\simeq30$ for $\tau^+\tau^-$. These are different tests. Diffuse-emission and unmodeled-source uncertainties prevent a straightforward Gaussian detection-significance interpretation. In the $\tau^+\tau^-$ scan, the maximum lies at the 10 GeV lower boundary; extending the scan downward did not find a stable interior maximum. This does not establish a preferred particle mass. Bin-by-bin spectra also do not distinguish the NFW and empirical two-point-source interpretations convincingly.

The authors searched gamma-ray, supernova-remnant, and pulsar catalogs without finding a compelling compact counterpart. The ROI overlaps the North Polar Spur/Loop~I radio structure, motivating attention to diffuse foregrounds, but the slowly varying 408 MHz emission is not a detailed morphological match to the residual. Neither that overlap nor the residual itself is independent confirmation of the sub-halo.

A complementary search for monochromatic lines was made by Salces P\'erez et al.\ \cite{Salces2026}, using Fermi-LAT data in the $10$--$300$ GeV range on the same candidate. They adopt the same fixed-NFW template ($r_s=0.1$ kpc) and obtain $J_{\rm ROI}=1.15\times10^{23}~\mathrm{GeV^2\,cm^{-5}}$, in agreement with Ref.~\cite{Zhu2025}. They find no statistically significant line; the largest excess, at about 28 GeV, has a local significance of $2.5\sigma$ and they do not claim it. Their limits on $\langle\sigma v\rangle_{\gamma\gamma}$ are stronger than existing Galactic Center line limits over much of the explored mass range, a comparison that, as the authors note, is driven by the large inferred $J$ rather than by a better analysis. They propagate the position uncertainty by scanning the centroid within $2\sigma$ and also note that tidal stripping could change the profile. A line search relies on a spectral feature rather than on morphology and is therefore less sensitive to the unmodeled continuum residual, but the conversion of a line flux into $\langle\sigma v\rangle$ still carries the same profile and $J$ uncertainty as the continuum analysis: the free-$r_s$ result of Sec.~\ref{sec:reference-params} applies to it equally. Because the same mass information enters both limits, they should not be combined as independent constraints.

Zhu et al.\ also quantify several background variations. Adding the two residual point sources changes the limits by factors of 1.0--1.4 ($b\bar b$) and 1.0--1.3 ($\tau^+\tau^-$). A Gaussian residual component with fitted width $\sigma\simeq1.26^\circ$ strengthens them by factors of 1.4--5 and 1.1--3, respectively. Changing the radius for free catalog-source parameters from $8^\circ$ to $5^\circ$, or replacing the Galactic diffuse template with a Planck-dust-opacity plus model-A IC construction, changes the limits by less than 0.5\% in each tested case. These are reported, mass-dependent tests within the fiducial halo setup, not an exhaustive foreground uncertainty budget. Retaining the residual is comparatively conservative among those tested descriptions, but does not by itself establish frequentist coverage for arbitrary foreground misspecification.

\subsection{Prompt annihilation and velocity-dependent rates}
\label{sec:indirect-annihilation}
The baseline in Sec.~\ref{sec:indirect-baseline} addresses specified annihilation channels and a fixed NFW shape. Its microscopic interpretation must be extended consistently when the annihilating fraction, velocity dependence, or halo evolution changes.

For self-conjugate DM with a spatially constant $s$-wave rate,
\begin{equation}
\frac{d\Phi_\gamma}{dE}=\frac{\langle\sigma v\rangle}{8\pi m_\chi^2}
J_\chi\frac{dN_\gamma}{dE}.
\label{eq:ann-flux}
\end{equation}
For a component tracing the total density at constant fraction $f_\chi$, $J_\chi=f_\chi^2J$. For asymmetric or non-self-conjugate populations, use the actual particle/antiparticle densities and corresponding normalization. A gamma-ray limit cannot be applied unchanged to every thermal component.

If $\sigma v=(\sigma v)_0K(v_{\rm rel})$, replace $J_\chi$ with
\begin{align}
J_K={}&\int d\Omega\,ds\,\rho_\chi^2(r)
\int d^3v_1d^3v_2\nonumber\\
&\times p_r(\mathbf v_1)p_r(\mathbf v_2)
K(|\mathbf v_1-\mathbf v_2|),
\label{eq:Jvelocity}
\end{align}
where each $p_r$ is normalized. At low velocities, $p$-wave annihilation is suppressed, while Sommerfeld enhancement can increase the rate until saturation, with possible resonances \cite{Boddy2017}. Density and velocity distributions both change during SIDM evolution or tidal stripping. A thermal-relic annihilation rate at freeze-out is not automatically the present rate in this target.

For an unchanged photon spectrum and normalized spatial template, a pure normalization change permits the conditional rescaling
\begin{equation}
\mu_{95}^{\rm new}=\mu_{95}^{\rm ref}
\frac{J_{\rm ref}}{f_\chi^2J_{\rm new}},
\label{eq:ann-recast}
\end{equation}
provided the nuisance treatment is transformed consistently and the self-conjugate convention is retained. This identity does not justify rescaling between different spatial profiles, velocity-weighted maps, or photon spectra: these require the counts likelihood. In particular, a collapsed SIDM cusp can change both $J_K$ and its degeneracy with localized residuals, while a core changes the overlap with diffuse emission. A thermal-relic comparison must use the present-day rate, branching fractions, and abundance of the specified model.

\subsection{Radiative decay}
\label{sec:indirect-decay}
For a decaying constituent,
\begin{equation}
\frac{d\Phi_\gamma}{dE}=
\frac{D_\chi}{4\pi\tau_\chi m_\chi}\frac{dN_\gamma}{dE},
\label{eq:decay-flux}
\end{equation}
with $D_\chi=f_\chi D$ only if the fraction is spatially constant. An assumed aperture-integrated flux threshold gives the diagnostic scale
\begin{align}
\tau_{\rm sens}&=\frac{D_\chi N_\gamma}{4\pi m_\chi\Phi_{\rm sens}}\nonumber\\
&\simeq7.6\times10^{27}~\mathrm{s}\, f_\chi
\left(\frac{D}{3.2\times10^{21}~\mathrm{GeV\,cm^{-2}}}\right)\nonumber\\
&\quad\times\left(\frac{N_\gamma}{3}\right)
\left(\frac{100~\mathrm{GeV}}{m_\chi}\right)
\left(\frac{10^{-9}~\mathrm{cm^{-2}\,s^{-1}}}{\Phi_{\rm sens}}\right).
\label{eq:decay-bound}
\end{align}
Here $N_\gamma$ refers to the chosen energy band. The nominal flux threshold is not a derived sensitivity for a $23^\circ$ source, so Eq.~\eqref{eq:decay-bound} is not an observed lifetime bound. A decay template is broader than the annihilation template and needs its own foreground analysis. Invisible recoil decays in Sec.~\ref{sec:invisible} have different branching ratios, lifetimes, and observables.

Zhu et al.\ do not derive a decay-lifetime constraint: their fitted template is proportional to $\rho^2$ and their particle scan concerns annihilation. Their LAT data selection and foreground study provide a starting point, but an annihilation curve cannot be converted into a decay bound by replacing $J$ with $D$ alone.

\subsection{Secondary synchrotron and inverse-Compton emission}
\label{sec:indirect-multiwavelength}
Injected electrons and positrons produce synchrotron and inverse-Compton (IC) radiation as they propagate \cite{ColafrancescoProfumoUllio2006,Storm2013}. The candidate's location near the Galactic disk motivates using local magnetic-field and radiation-field information, but does not eliminate their spatial uncertainty or the much brighter diffuse foreground. As illustrative environmental values take $B=3$--$6~\mu\mathrm G$, $U_{\rm ISRF}\sim1~\mathrm{eV\,cm^{-3}}$, and $U_{\rm CMB}=0.26~\mathrm{eV\,cm^{-3}}$.

For pitch angle of order unity the synchrotron critical frequency is
\begin{equation}
\nu_c\simeq16~\mathrm{MHz}
\left(\frac{B}{1~\mu\mathrm G}\right)
\left(\frac{E_e}{1~\mathrm{GeV}}\right)^2.
\label{eq:nu-synch}
\end{equation}
The single-electron spectral maximum is near $0.29\nu_c$, and an injected spectrum and pitch-angle average broaden it. At $B=3~\mu\mathrm G$ and $E_e=10$--50 GeV, $\nu_c\simeq5$--120 GHz; lower-frequency surveys mainly probe other portions of the evolved electron distribution. IC photon energies depend on both the electron and the target-photon energies and extend from the X-ray band into the GeV range.

In the Thomson limit the cooling time is
\begin{equation}
t_{\rm cool}=\frac{3m_e^2c^3}
{4\sigma_T E_e(U_B+U_{\rm rad})},\qquad U_B=B^2/(8\pi).
\label{eq:tcool}
\end{equation}
For $B=3~\mu\mathrm G$ and $U_{\rm rad}=1.26~\mathrm{eV\,cm^{-3}}$, the total energy density is about $1.48~\mathrm{eV\,cm^{-3}}$, giving $t_{\rm cool}\simeq21$ Myr at 10 GeV and 4.2 Myr at 50 GeV before Klein--Nishina corrections. As an example only, $D_{\rm diff}=3\times10^{28}~\mathrm{cm^2\,s^{-1}}$ gives $r_s^2/(6D_{\rm diff})\simeq1.7\times10^4$ yr. With these assumptions the central $0.1$ kpc region is not calorimetric: emission can be spatially much broader than the injection profile. Escape from the sub-halo does not mean escape from the Galactic diffusion region.

The 408 MHz map and IC foreground used in Ref.~\cite{Zhu2025} characterize astrophysical backgrounds; they are not a calculation or limit on sub-halo-induced synchrotron or propagated IC emission. Such secondary components can differ substantially from its prompt-annihilation template and must be modeled separately.

\subsection{Shared uncertainties and interpretation}
\label{sec:indirect-caveats}
The emission and gravitational data can be combined through a common halo posterior, but the conditional large $J$ factor is not evidence for a halo. Gamma-ray residuals, candidate reliability, morphology, distance, and particle fractions all remain nuisance parameters or competing hypotheses. The framework applies to future secure objects after refitting their parameters; its numerical values do not transfer unchanged to another source. A future joint dynamical--LAT analysis should use the underlying photon likelihood and common structural parameters, rather than multiply a new dynamical constraint by a LAT limit that already incorporated the same mass information through its $J$ nuisance term.

\section{A conditional connection to the LZ recoil candidate}
\label{sec:lz}
\subsection{An inelastic nuclear-scattering interpretation}
LZ reports a single event consistent with a nuclear recoil of $248\pm23~\mathrm{(stat)}\pm23~\mathrm{(sys)}$ keV in an extended recoil window, with a maximum local significance of $3.4\sigma$ and a global significance of $2.6\sigma$ in its tested hypotheses \cite{LZ2026}. This is an event of interest, not an established DM detection. It neither confirms the pulsar perturber nor associates a particle with its direction.

One proposed interpretation is endothermic scattering $\chi_1N\to\chi_2N$, with $\delta=m_2-m_1>0$; TeV-scale Higgsinos with several-hundred-keV splittings provide examples \cite{FanReece2026,Freese2026}. With $c=1$ in kinematic expressions,
\begin{equation}
v_{\min}(E_R)=\frac{m_NE_R/\mu_{\chi N}+\delta}
{\sqrt{2m_NE_R}},
\label{eq:inelastic-vmin}
\end{equation}
so the interpretation depends strongly on the high-speed tail and nuclear interaction. It does not establish a long-lived cosmological population of $\chi_2$, its dark-sector self-scattering rate, or an invisible branching ratio.

\subsection{Decay and downscattering have different kick scales}
\label{sec:lz-kicks}
For an invisible transition $\chi_2\to\chi_1+\phi$ with massless $\phi$ and $\delta\ll m_\chi$, Eq.~\eqref{eq:decay-kick} gives
\begin{equation}
v_k\simeq0.105~\kms
\left(\frac{\delta}{350~\mathrm{keV}}\right)
\left(\frac{1~\mathrm{TeV}}{m_\chi}\right).
\label{eq:lz-smallkick}
\end{equation}
The same splitting in this channel is too small to appreciably heat the reference halo. A 10--30 $\kms$ decay kick at 1 TeV instead requires a splitting of approximately 30--100 MeV. A massive radiation quantum or additional daughters would change these kinematics.

By contrast, in $\chi_2\chi_2\to\chi_1\chi_1$, neglecting initial motion, each massive daughter receives kinetic energy approximately $\delta$. Its speed is
\begin{align}
v_{\rm out}&\simeq c\sqrt{\frac{2\delta}{m_\chi}}\nonumber\\
&\simeq250~\kms
\left(\frac{\delta}{350~\mathrm{keV}}\right)^{1/2}
\left(\frac{1~\mathrm{TeV}}{m_\chi}\right)^{1/2}.
\label{eq:lz-downscatter}
\end{align}
Such daughters could escape the reference sub-halo. Nuclear upscattering could constrain a splitting and portal interaction, while sub-halo survival could constrain the excited fraction and de-excitation rate. This is conditional complementarity, not a universal prediction of the Higgsino interpretation. If the excited population is depleted early or the self-interaction is negligible, survival supplies no useful test of this mechanism.

\subsection{Boosted daughters from a nearby source}
A different possibility is an energetic daughter population generated by parent decay or annihilation. Liang et al.\ explore a boosted interpretation of the LZ event with a momentum-dependent magnetic dipole--dipole interaction; their concrete model uses halo annihilation into mediators followed by mediator decay, not the slow parent-decay scenario of Sec.~\ref{sec:invisible} \cite{Liang2026}. A viable spectrum must account for the high-energy event without producing an incompatible lower-energy population; normalization to one event alone is not validation.

For freely propagating daughters from a steady, decaying source with current parent-density factor $D_X$, the all-source spectral number flux is
\begin{equation}
\frac{d\Phi_Y}{dE_Y}=\frac{D_X}{4\pi m_X\tau_X}
\frac{dN_Y}{dE_Y}.
\label{eq:boost-flux}
\end{equation}
Use the appropriate travel time and source evolution if steady state fails. If the present parent fraction traces total mass, $D_X=f_XD$; for escaping daughters this assumption describes the parents, not the daughter density at Earth. A local annihilation source instead has a density-squared dependence, modified by mediator propagation if its decay length is significant.

The predicted event count requires a detector convolution,
\begin{align}
N_{\rm evt}={}&\mathcal{E}\sum_T N_T^{(1)}\int dE_Y\,
\frac{d\Phi_Y}{dE_Y}\int dE_R\,
\frac{d\sigma_{YT}}{dE_R}\nonumber\\
&\times\int_{\rm window}dE_{\rm obs}\,
\epsilon_{\rm det}(E_{\rm obs})
\mathcal{R}(E_{\rm obs}\mid E_R),
\label{eq:boost-count}
\end{align}
where $\mathcal E$ is mass-time exposure, $N_T^{(1)}$ the number of target nuclei per detector mass, and $\mathcal R$ the energy-response density. For elastic scattering on xenon,
\begin{equation}
v_{\min}=\sqrt{\frac{m_NE_R}{2\mu_{YN}^2}},\qquad
\mu_{YN}v_{\min}\simeq0.123~\mathrm{GeV}
\label{eq:lz-elastic-threshold}
\end{equation}
at $E_R=248$ keV and $m_N\simeq122$ GeV. Even the very-heavy-projectile limit requires about 300 $\kms$ in the detector frame. A tens-of-$\kms$ kick is therefore not by itself the energetic boost needed for this event, although all incident speeds must include source motion and Galactic propagation.

For the truncated reference the Sun is outside the bound source. Its large central density cannot be substituted for the local density in an ordinary halo-scattering rate. Escaping daughters could in principle reach Earth, but the smooth Milky Way and cosmological source populations generally also contribute and must be compared. For freely streaming daughters the all-source flux is proportional to the whole-source $D$ (parent decay) or $J$ (parent annihilation) divided by $4\pi$, so the comparison with the smooth halo reduces to $D_{\rm sub}/D_{\rm MW}$ or $J_{\rm sub}/J_{\rm MW}$ at the same abundance and rates, with the Milky Way integrated over all directions. For an NFW Milky Way with $r_s=20$ kpc and $0.3~\mathrm{GeV\,cm^{-3}}$ at $R_\odot=8.122$ kpc, $D_{\rm MW}\simeq2.0\times10^{23}~\mathrm{GeV\,cm^{-2}}$ and $J_{\rm MW}\simeq8.3\times10^{22}~\mathrm{GeV^2\,cm^{-5}}$, against $3.2\times10^{21}$ and $1.2\times10^{23}$ for the truncated sub-halo of Eq.~\eqref{eq:JD-values}. A decaying parent in the sub-halo therefore supplies only about $1.6\%$ of the Galactic flux and cannot be the dominant source. The annihilation case differs: the sub-halo's total $J$ is comparable to that of the entire Galactic halo (a ratio of 1.5 here; 0.9--2.5 for generalized-NFW inner slopes $\gamma=0$--1.2, and 0.1 for $\gamma=1.5$) and exceeds that of the central 1 and 3 kpc of the Galaxy ($1.2\times10^{22}$ and $3.2\times10^{22}~\mathrm{GeV^2\,cm^{-5}}$) by factors of about 10 and 4. Hence, if in a given annihilation-driven model the Galactic center and smooth halo cannot supply the flux required to explain the event, a nearby dense sub-halo would be a competitive and possibly dominant source; if they can, the sub-halo is not needed. This is a scale comparison, not a prediction: it uses the unrefitted fixed-NFW cusp, whose $J$ varies by a factor of about 75 among the profiles of Sec.~\ref{sec:future-emission}, and ignores mediator propagation and anisotropy. A nondirectional nuclear recoil cannot establish spatial association with this particular sub-halo. Large decay kicks can remove mass from the source, so a dominant rapidly decaying parent needs a simultaneous survival calculation; a long lifetime or subdominant source fraction can change that conclusion but reduces the flux.

\section{Discriminating among dark matter models with new data}
\label{sec:future}
Sections~\ref{sec:fdm}--\ref{sec:lz} show that, if the candidate is a dark sub-halo, nearly every microphysical question it raises turns on quantities that the present data do not measure. We now ask which new measurements would separate the model classes. No new data analysis is performed; the numbers below illustrate scales with the reference profile of Table~\ref{tab:reference} and the illustrative profiles of Fig.~\ref{fig:obs2-identifiability}, and they are not forecasts from a completed sensitivity study.

The models differ mainly through four quantities: (i) whether the perturber is real, extended, and dark; (ii) its interior mass distribution on scales of roughly 10--300 pc; (iii) its velocity relative to the disk and to the host halo; and (iv) the abundance of comparable objects in a volume with known selection. The existing endpoint accelerations constrain none of (ii)--(iv). Every nominal endpoint lies outside the illustrative tidal radius, where the shell theorem removes all dependence on interior structure (Sec.~\ref{sec:reference} and Fig.~\ref{fig:obs2-identifiability}); the velocity enters only through a centroid drift far below the positional uncertainty (Sec.~\ref{sec:future-motion}); and a single candidate without a calibrated selection function supplies no population. Table~\ref{tab:discriminants} lists, for each model class, the signature that would distinguish it from a tidally truncated CDM halo of the same bound mass and the new observable most directly sensitive to it. We discuss four routes in turn: accelerations measured inside the halo (Sec.~\ref{sec:future-accel}), lensing and accretion as tests of compactness (Sec.~\ref{sec:future-compact}), the morphology of any annihilation or decay signal (Sec.~\ref{sec:future-emission}), and a search for bound stars (Sec.~\ref{sec:future-dark}). No single route measures both the interior profile and the velocity for every model class, which is the main limitation of this program.

\begin{table*}[t]
\caption{Signatures that distinguish each model class from a tidally truncated CDM halo of the same bound mass, and the new observables most directly sensitive to them. Numbers refer to the reference profile of Table~\ref{tab:reference} or the illustrative profiles of Sec.~\ref{sec:future-accel}, at $d=0.78$ kpc and, where a speed enters, $V=250~\kms$.}
\label{tab:discriminants}
\begin{ruledtabular}
\begin{tabular}{p{0.19\textwidth}p{0.35\textwidth}p{0.38\textwidth}}
Model class & Distinguishing signature & Most direct new-data test \\
\hline
Single compact object (black hole, compact cluster) & All mass within $\lesssim10$ pc; Einstein radius $\simeq9$--$16''$; Bondi accretion from the interstellar medium & Interior $a\propto r^{-2}$; astrometric anomalies $\gtrsim1$ mas yr$^{-1}$ within $\sim2'$; radio, millimeter, and X-ray counterpart; no annihilation signal \\
Truncated CDM & Cusp: $d\ln a/d\ln r\simeq-0.3$ at 30 pc to $-1$ at 200 pc; $J$ dominated by the center ($\simeq40\%$ within $1^\circ$) & Accelerations at two interior radii; annihilation morphology; population with calibrated selection \\
WDM near half-mode & Lower concentration at fixed mass; suppressed abundance & Interior $a(r)$ after a stripping model; population counts \\
Wave DM & Soliton core $r_c\propto m^{-1}M^{-1/3}$; granules on $\lambda_{\rm dB}$ & Unresolved by accelerometers for allowed $m$ ($r_c\lesssim20$ pc, $\lambda_{\rm dB}\approx6$ pc or less); inner annihilation morphology at $1^\circ\simeq14$ pc if a signal exists; PTA signals \\
SIDM, expanded core & Core of order $r_s$; $a\propto r$ inside it; low $J$ at fixed $D$ & Rising $a(r)$ at interior radii; $J$ several times below the cusp value \\
SIDM, collapsed & Steeper-than-NFW center inside a retained envelope & Steep inner $a(r)$ with an extended outer $M(r)$, unlike a compact object; point-like plus extended annihilation \\
Invisible decay & Expanded, lower-density profile and mass loss & Interior profile plus orbit to separate from tidal heating; population of central densities \\
PBH-composed halo & Stellar-mass constituents; collisional core or mass segregation & Microlensing optical depth $\simeq6\times10^{-8}f_c$ near 50 pc, $t_E\simeq12$ d $(M_c/\msun)^{1/2}$, lenses sharing one proper motion \\
Dissipative component & Flattened or disk-aligned; possibly low $V$ & Anisotropic interior accelerations; low $V$ from any bound tracers \\
Luminous alternative (faint dwarf) & Bound stars & Comoving Gaia clump at $\varpi\simeq1.3$ mas, $|\mu|\simeq68$ mas yr$^{-1}$ $(V_\perp/250~\kms)$ \\
Annihilating or decaying DM & Emission tracing $\rho^2$ or $\rho$ & Angular profile of the emission; $J/D$, which varies by $\sim75$ at fixed $M_t$ \\
\end{tabular}
\end{ruledtabular}
\end{table*}

\begin{figure*}[t]
\centering
\includegraphics[width=\textwidth]{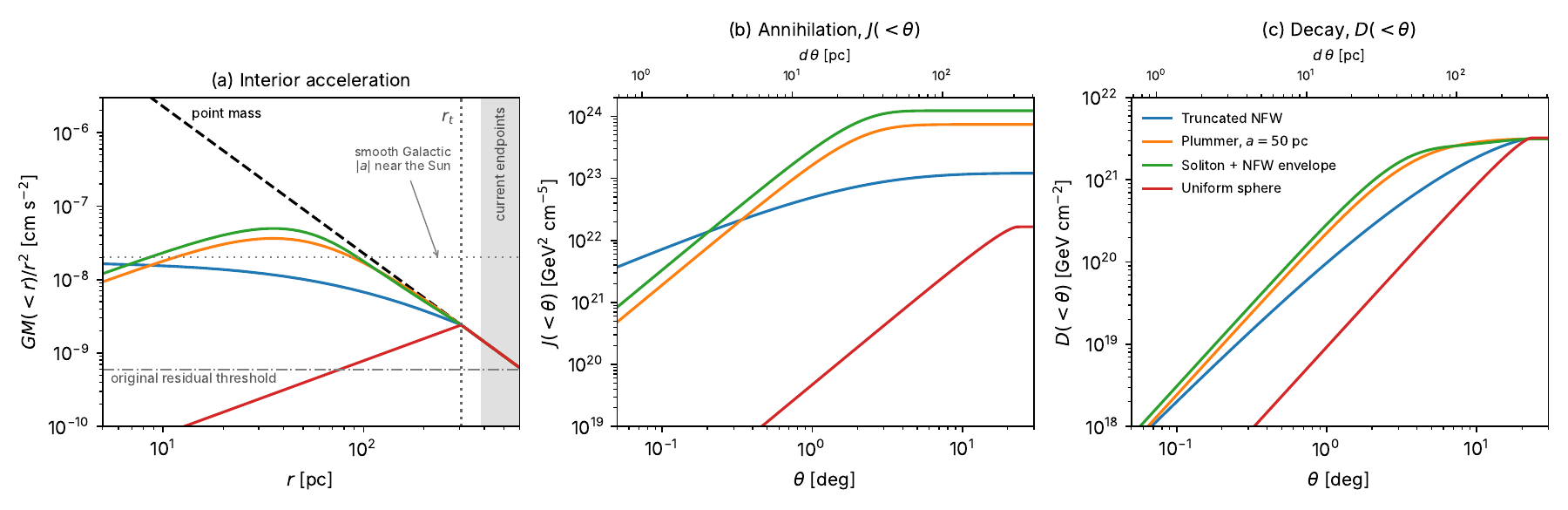}
\caption{Observables that separate interior profiles which the current endpoint accelerations cannot. All profiles share the bound mass $M_t=1.63\times10^7\msun$ inside $r_t=0.307$ kpc at $d=0.78$ kpc, as in Fig.~\ref{fig:obs2-identifiability}. (a) Unprojected acceleration magnitude $GM(<r)/r^2$; the curves coincide outside $r_t$, where every current endpoint lies (shaded, beginning at the nearest nominal endpoint), and differ by up to two orders of magnitude inside it. The dash-dotted line is the $6\times10^{-10}~\mathrm{cm\,s^{-2}}$ residual threshold of the original search and the dotted line the approximate smooth Galactic acceleration near the Sun. (b) Cumulative annihilation factor and (c) cumulative decay factor within angle $\theta$ of the center, from the exact line-of-sight integral; the top axes give the corresponding projected radius. The point mass (dashed) appears only in (a), since a compact object supplies no particle-annihilation or decay template; the legend in (c) identifies the extended profiles in all three panels.}
\label{fig:future-discriminants}
\end{figure*}

\subsection{Accelerations inside the halo}
\label{sec:future-accel}
\paragraph{Interior endpoints and the radial scaling.}
Pulsar accelerations become sensitive to the profile only when endpoints fall inside the halo: a single timed source inside $r_t$ measures an interior $M(<r)$ directly, whereas any number of exterior endpoints measure only $M_t$. The discriminating power then comes from the radial scaling of $a=GM(<r)/r^2$, which falls as $r^{-2}$ for a point mass, is nearly flat for an NFW cusp ($M\propto r^2$), and rises as $r$ inside a core. For a point mass, the truncated NFW profile, and the Plummer core of Fig.~\ref{fig:obs2-identifiability}, and for a constant-density sphere filling $r_t$, all normalized to $M_t=1.63\times10^7\msun$, the unprojected magnitudes at $r=50$, 100, and 200 pc are $(9.1,2.3,0.57)\times10^{-8}~\mathrm{cm\,s^{-2}}$ for a point mass, $(1.0,0.68,0.38)\times10^{-8}$ for the truncated NFW profile, $(3.3,1.7,0.54)\times10^{-8}$ for a Plummer core with $a=50$ pc, and $(0.04,0.08,0.16)\times10^{-8}$ for a constant-density sphere filling $r_t$. Figure~\ref{fig:future-discriminants}(a) shows the full curves. The local logarithmic slope $d\ln a/d\ln r$ is $-2$, $-0.3$ to $-1$, $+0.2$ to $-1.8$, and $+1$ between 30 and 200 pc in the same four cases. Except deep inside a large core, these signals are a substantial fraction of, or larger than, the smooth Galactic acceleration at the Solar circle, about $2\times10^{-8}~\mathrm{cm\,s^{-2}}$, and one to two orders of magnitude above the $6\times10^{-10}~\mathrm{cm\,s^{-2}}$ residual threshold of the original search. Two interior endpoints at different radii would therefore separate the compact, cuspy, and cored branches, and the limitation is the availability of tracers inside $r_t$, not precision. Only the line-of-sight component is measured, so the predicted sign flips across the centroid and the Solar term of Eq.~\eqref{eq:obs2-forward} must be modeled; interior endpoints also constrain the centroid and so address the current degeneracy between the mass of the sub-halo and its position, including the position-box problem of the original fit (Appendix~\ref{app:obs2}).

\paragraph{Tracers.}
The present catalog has no endpoint inside $r_t$ (Table~\ref{tab:obs2-radii}). New millisecond pulsars from MeerKAT, FAST, and eventually SKA surveys are therefore valuable mainly through their positions relative to the centroid, and longer baselines improve the binary-period derivatives that already exist. The precision of the observed period drift rate improves as $T^{-5/2}$ with the baseline $T$, while the proper motion and parallax uncertainties scale as $T^{-3/2}$ and $T^{-1/2}$, respectively \cite{LorimerKramer2005}. Donlon et al.\ \cite{Donlon_Chakrabarti_Donghia2026} discuss the gain of direct acceleration measurements over kinematic estimates and find that, if the current trend continues, about 200 direct acceleration measurements could determine the local dark matter density at better than $25\sigma$. Other direct accelerometers, such as extreme-precision radial-velocity monitoring \cite{Chakrabarti2020,Chakrabarti_Wagner2026} and eclipse timing of detached binaries \cite{Chakrabarti2022}, could add interior endpoints wherever suitable stars lie inside $r_t$, since their targets are far more numerous than pulsars; a signal of $10^{-8}~\mathrm{cm\,s^{-2}}$ corresponds to a radial-velocity drift of about $3~\mathrm{cm\,s^{-1}}$ per decade. The eclipsing binaries identified in the \textit{Kepler} field by Ref.~\cite{Chakrabarti2022}, for which other dynamical effects are subdominant to the Galactic signal, are already more numerous than known millisecond pulsars. The \textit{Kepler} field lies about $45^\circ$ from the direction of the candidate, however, so that no \textit{Kepler} sightline passes within about 0.5 kpc of its centroid: these binaries add exterior endpoints rather than tracers inside $r_t$, and interior tracers require targets selected toward the candidate. Ongoing timing of eclipsing double white dwarfs in the Solar neighborhood may provide another means of finding tracers within $r_t$ (S.~Chakrabarti et al., in preparation). Of particular relevance is a future measurement of the \emph{radial variation} of the local dark matter density, for which Donlon et al.\ \cite{DonlonWidrow2025} developed a framework that applies Gauss's law to line-of-sight accelerations on spheres about the Sun and constrains the enclosed mean density without a parametric potential. A measurement of that variation is not yet possible: the current sensitivity of the method reaches only $\gtrsim10^8\msun$ objects, so it tests the population rather than this candidate's profile, and it is less sensitive to dark matter substructure than individual direct acceleration measurements \cite{Chakrabarti2021}.

\paragraph{Interpreting a profile.}
A generalized profile,
\begin{equation}
\rho(r)=\frac{\rho_s}{(r/r_s)^\gamma[1+r/r_s]^{3-\gamma}},
\label{eq:zhao}
\end{equation}
can organize forecasts, provided it is truncated and not forced on solitons or compact remnants. A flat inner profile is not unique to SIDM or wave DM, and a dense cusp is not unique to collisionless CDM. For SIDM, a measured profile must be mapped to the evolutionary branches of Sec.~\ref{sec:sidm-collapse}, and a steep center inside an extended outer $M(r)$ distinguishes a collapsed halo from a compact object, which has no envelope. For wave DM, the soliton shape, density--radius relation, and envelope test consistency; with the core--halo relation of Eq.~\eqref{eq:soliton}, $m\geq2\times10^{-20}$ eV implies $r_c\lesssim20$ pc for this mass, and the internal de Broglie wavelength at $v\simeq15~\kms$ is at most about 6 pc, so resolving either would require tracers within about 20 pc of the center. Concentration at fixed bound mass separates WDM near its half-mode scale from CDM \cite{Lovell2014} only after the tidal stripping that also lowers concentration has been modeled; one object gives weak evidence. Expanded profiles from invisible decays face the same degeneracy with tidal heating, which an orbit would partly break.

For fermionic DM with dimensionless primordial maximum occupation $f_{\max}$ per internal state, a Maxwellian coarse-grained distribution obeys
\begin{equation}
\rho\leq\frac{g f_{\max}m_f^4}{h_P^3}
(2\pi\sigma_{\rm 1D}^2)^{3/2},
\label{eq:tremaine-gunn}
\end{equation}
where $h_P$ is Planck's constant and $g$ the degeneracy \cite{TremaineGunn1979}. For $g f_{\max}=1$, $m_f=6$ keV, and an assumed $\sigma_{\rm 1D}=15~\kms$, the right-hand side is about $3.5\times10^4\msun\,\mathrm{pc^{-3}}$, much larger than the reference density near $r_s$. The bound is therefore not competitive at any radius that accelerometers can sample, and the thermal-equivalent WDM mass is not automatically the physical fermion mass entering it.

\subsection{Compactness: lensing and accretion}
\label{sec:future-compact}
\paragraph{An extended halo is not a usable lens.}
For extragalactic sources the critical surface density at $d=0.78$ kpc is $c^2/(4\pi Gd)\simeq2\times10^9\msun\,\mathrm{pc^{-2}}$, so the reference profile, with $\Sigma\simeq260\msun\,\mathrm{pc^{-2}}$ at $R=50$ pc, has convergence of order $10^{-7}$. Astrometric weak lensing of background sources \cite{VanTilburg2018,Mondino2020} induces proper motions of order $10^{-5}$ mas yr$^{-1}$ or less for this extended, slowly moving target. Lensing by the smooth halo as a whole is therefore not a practical observable for current facilities. Lensing instead tests the two branches in which the mass is compact.

\paragraph{A single compact object.}
A black hole or compact cluster with $M_t\simeq(1.6$--$2.5)\times10^7\msun$ at 0.78 kpc is a strong lens. Its Einstein radius is $\theta_E\simeq9$--$11''$ for sources at twice its distance and $13$--$16''$ for distant sources. With a relative proper motion of $V_\perp/d\simeq68$ mas yr$^{-1}$ $(V_\perp/250~\kms)$, the induced astrometric motion of background stars, $\simeq\theta_E^2\mu/\theta^2$ for $\theta\gg\theta_E$, exceeds 1 mas yr$^{-1}$ within roughly 1.3--2.2 arcmin, accompanied by photometric magnification of sources within a few $\theta_E$. The pattern is small but strong, and a template search over the region allowed by the acceleration fit, using Gaia proper motions and epoch astrometry, would find or exclude it; an extended halo predicts nothing comparable. The same object moving through the interstellar medium accretes. Its Bondi radius $2GM/V^2$ is about 3 pc at $V=250~\kms$, and for the compact-fit mass the Bondi rate gives $\dot Mc^2\sim5\times10^{40}$--$5\times10^{42}~\mathrm{erg\,s^{-1}}$ for ambient densities of 0.003--0.3 cm$^{-3}$ \cite{Bondi1952}, scaling as $V^{-3}$. Even at a radiative efficiency of $10^{-6}$ relative to $\dot Mc^2$, the flux at Earth would be $\sim10^{-9}$--$10^{-7}~\mathrm{erg\,cm^{-2}\,s^{-1}}$. Low-rate accretion flows are inefficient and the Bondi rate overestimates the inflow, so this is not an exclusion. However, comparing a calculated spectrum with existing radio, millimeter, and X-ray catalogs over the allowed region requires no new data and is the most immediate test of the single-object branch. The catalog searches reported in Ref.~\cite{Zhu2025} concerned gamma-ray sources, supernova remnants, and pulsars.

\paragraph{A halo of stellar-mass constituents.}
If a fraction $f_c$ of the halo mass is in compact constituents of mass $M_c$, the microlensing optical depth for sources at $D_s$ behind it is \cite{Paczynski1986}
\begin{equation}
\tau(\mathbf R)=f_c\,\frac{4\pi G}{c^2}\,\Sigma(\mathbf R)\,d\left(1-\frac{d}{D_s}\right).
\label{eq:tau-halo}
\end{equation}
For the reference profile and $D_s=2d$, $\tau\simeq6\times10^{-8}f_c$ at $R=50$ pc, $3\times10^{-8}f_c$ at 100 pc, and $5\times10^{-8}f_c$ averaged within $r_s$ (about $7^\circ$ in radius); the values double for distant sources. Event timescales are $t_E\simeq12~\mathrm{d}\,(M_c/\msun)^{1/2}(250~\kms/V_\perp)$, with $\theta_E\simeq2.3$ mas $(M_c/\msun)^{1/2}$, and the rate per monitored source is $(2/\pi)\tau/t_E\simeq10^{-6}f_c(M_c/\msun)^{-1/2}~\mathrm{yr^{-1}}$. A few events per year therefore require monitoring of order $10^6$--$10^7$ sources behind the central few degrees, a scale within reach of wide-field time-domain surveys \cite{Ivezic2019} but requiring a model of the source population at 1--3 kpc. Three features distinguish a constituent signal from disk self-lensing: the optical depth is concentrated at the dynamical centroid and traces $\Sigma(\mathbf R)$; all lenses share the halo's proper motion, so the relative proper motions of events align; and astrometric microlensing can measure individual lens masses and proper motions, as demonstrated for an isolated stellar-mass black hole \cite{Sahu2022}. A detection would therefore measure $M_c$, the projected profile, and the halo's transverse velocity together. Particle dark matter predicts no excess, so a null result in a well-modeled survey bounds $f_c(M_c)$ for this object, conditional on its mass and profile. This route complements the collisional diagnostics of Sec.~\ref{sec:pbh}, which constrain $M_c$ only through the profile.

\subsection{Profile-sensitive emission}
\label{sec:future-emission}
Gamma-ray and laboratory data discriminate among particle models only once the structure is fixed, but emission morphology can also help fix it. The decay factor depends only on the enclosed mass: for the extended profiles above at fixed $M_t$, the whole-source $D$ is $(3.16$--$3.24)\times10^{21}~\mathrm{GeV\,cm^{-2}}$. The annihilation factor does not. With the exact integral of Eq.~\eqref{eq:exact-spherical}, $J_{\rm all}=1.2\times10^{23}~\mathrm{GeV^2\,cm^{-5}}$ for the truncated NFW profile, $7.4\times10^{23}$ for the Plummer core with $a=50$ pc, $1.2\times10^{24}$ for the soliton-plus-envelope profile of Fig.~\ref{fig:obs2-identifiability}, and $1.7\times10^{22}$ for a constant-density sphere filling $r_t$, a range of about 75 at fixed mass [Fig.~\ref{fig:future-discriminants}(b,c)]. The fraction of $J$ within $1^\circ$, $2^\circ$, and $5^\circ$ of the center is $(0.41,0.61,0.86)$ for the NFW profile, $(0.22,0.60,0.97)$ for the Plummer core, and $(0.00,0.01,0.07)$ for the uniform sphere. At $0.78$ kpc, $1^\circ$ is about 14 pc, so the angular profile of a detected annihilation signal resolves the interior on scales the present accelerometers cannot reach, and a collapsed SIDM core would appear as a point-like component on extended emission. A compact object gives no annihilation signal at all. If a particle model produces both channels, $J/D$ fixes a density scale independently of the mass normalization. With a profile measured by any of these routes, the annihilation and decay likelihoods of Sec.~\ref{sec:indirect} can be refitted with templates that are no longer assumptions, and the relative strength of $\rho^2$, $\rho$, and transported secondary emission becomes a test of the annihilating or decaying fraction rather than of the halo model. Velocity-dependent annihilation additionally needs the internal velocity distribution implied by the profile (Eq.~\eqref{eq:Jvelocity}). The degree-scale extent of a decay signal favors wide-field gamma-ray and X-ray instruments over narrow-field ones. The localized gamma-ray feature about $1.5^\circ$, or about 20 pc, from the dynamical centroid \cite{Zhu2025} can then be tested for co-location with a profile-informed template rather than with the fixed NFW one.

\subsection{Is it dark? Bound stars}
\label{sec:future-dark}
A dark sub-halo and an undiscovered faint dwarf galaxy make the same pulsar predictions. Stars bound to the object would share its parallax, $\varpi\simeq1.3$ mas, and its proper motion relative to the Sun, $|\mu|\simeq68$ mas yr$^{-1}$ $(V_\perp/250~\kms)$, with an internal spread $\lesssim v_c/d\simeq4$ mas yr$^{-1}$. A coherent clump at large proper motion stands out against disk stars at the same distance. At a distance modulus of about 9.5, Gaia's magnitude limit reaches old main-sequence stars well below the turnoff, so a population search requires no new survey. A null result bounds the stellar mass and supports the dark interpretation; a detection would measure the object's velocity and centroid directly from its members, and their metallicities and ages would test a dwarf-galaxy origin. Gas and stellar mass from three-dimensional dust and star-count maps should likewise be compared with $M_t$ before the remainder is called dark; a giant molecular cloud (GMC) can produce comparable local accelerations but not a dark bound mass of $\sim10^7\msun$. Reference~\cite{Chakrabarti2025} searched GMC catalogs, atomic hydrogen catalogs, and Gaia data and concluded that the baryonic budget (molecular gas, atomic hydrogen, and stars) near the candidate sub-halo would not produce the measured acceleration.

\subsection{Motion, wakes, and pulsar-timing arrays}
\label{sec:future-motion}
\paragraph{Motion and assembly history.}
Without a luminous counterpart, pulsar timing could constrain a transverse velocity only through a time-varying acceleration pattern or a drifting fitted centroid. At $d=0.78$ kpc, $v_\perp=200~\kms$ corresponds to about 54 mas yr$^{-1}$, or only $0.4$ arcsec over eight years, vastly smaller than the current centroid uncertainty. Among the routes above, only bound stars (Sec.~\ref{sec:future-dark}) or the lens kinematics of the compact branches (Sec.~\ref{sec:future-compact}) supply $\mathbf V$; for an extended halo of particle dark matter without a stellar component, the velocity remains unmeasured by them. Even full present-day phase space does not uniquely determine infall time, because the host grows, the orbit and sub-halo mass evolve, and encounters can matter. Integrating backward in a fixed potential is not a measurement of assembly age; an orbit constrains pericenter and exposure to disk shocks, and must be combined with cosmological priors on the infall history, with internal SIDM evolution starting before infall. The velocity entering internal heat transport is the pairwise internal distribution, host evaporation uses the host--sub-halo relative distribution (Eq.~\eqref{eq:host-evap}), and direct detection uses the incident distribution in the laboratory; these should not be interchanged. A nondirectional nuclear-recoil detection cannot be associated with this object (Sec.~\ref{sec:lz}); a measured $\mathbf V$ and distance would fix the arrival direction and kinematics that any future directional signal from escaping daughters would need to match.

\paragraph{Dark matter wakes.}
The perturber raises a wake in the ambient dark matter. A classical order-of-magnitude deceleration is $a_{\rm df}\sim4\pi G^2M\rho\ln\Lambda/v^2$, with an additional velocity-distribution factor; for $M=2.45\times10^7\msun$, $\rho=0.3~\mathrm{GeV\,cm^{-3}}$, $v=250~\kms$, and $\ln\Lambda=5$ it is of order $10^{-11}~\mathrm{cm\,s^{-2}}$. This is a deceleration of the perturber, not a pulsar signal; comparing it with the acceleration threshold requires the wake field at the pulsar and Solar positions. Wave-DM corrections depend on dimensionless combinations of mass, speed, and length scales rather than on a universal boson-mass threshold \cite{Lancaster2020}.

\paragraph{Timing arrays and catalogs.}
Longer baselines and additional precisely timed pulsars improve sensitivity only in proportion to their spatial sampling and systematics. The coherent and stochastic pulsar-timing signatures of ultralight DM (Sec.~\ref{sec:fdm-granularity}) probe the ambient field rather than the sub-halo and remain an independent test of the same boson-mass range. A forecast for a specific future release (an IPTA combination, a MeerKAT or FAST sample, or a defined SKA scenario) requires the selection machinery of Sec.~\ref{sec:likelihood}. The relation to PTA searches for transiting substructure, which use the timing residuals rather than the accelerations, is treated in Sec.~\ref{sec:future-ptas}.

\subsection{Relation to timing-residual searches for transiting substructure}
\label{sec:future-ptas}
A substantial literature, largely from Zurek and collaborators, asks a question that resembles ours but is distinct: can the timing residuals of a PTA reveal dark matter substructure transiting near the Earth--pulsar system? We summarize what these works establish, what they imply for the candidate, and where the two programs meet. The values quoted from those papers are taken from the cited works; the estimates specific to the candidate are order-of-magnitude calculations, made with their expressions, and not new fits.

\paragraph{Framework and forecasts.}
Baghram, Afshordi, and Zurek \cite{Baghram2011} first considered the Doppler frequency shift that substructure imprints on pulsar timing. Dror, Ramani, Trickle, and Zurek \cite{Dror2019} studied single transiting compact objects, primordial black holes and compact sub-halos with masses from $10^{-12}\msun$ to well above $100\msun$, through the Doppler shift produced by the acceleration of the pulsar or the Earth and through the Shapiro delay of pulses passing near the line of sight. They found that the Square Kilometre Array could constrain such objects to be a subdominant component of the dark matter over this range, reaching the sub-percent level over parts of it, and that PTAs probe substantially less dense objects than lensing because of the large effective radius over which an object can be observed. Ramani, Trickle, and Zurek \cite{Ramani2020} generalized the treatment to the pulsar-phase correlations induced by general substructure, with projections for monochromatic primordial black holes and for CDM-like NFW sub-halos. Lee, Mitridate, Trickle, and Zurek \cite{LeeMitridate2021} extended it to diffuse sub-halos and to the small-scale power spectrum, including the effect of tidal disruption. Lee, Taylor, Trickle, and Zurek \cite{LeeTaylor2021} developed a Bayesian search with mock timing data and concluded that the main gap between idealized forecasts and realistic data is how well a gravitational-wave background (GWB) can be separated from the substructure signal.

That gap is the subject of the most recent work. The evidence for a nanohertz GWB \cite{NANOGrav2023} acts as red noise, and Refs.~\cite{Cherukupalli2026,Foster2026} show that it degrades the sensitivity to substructure by one to three orders of magnitude relative to white-noise forecasts. The dynamic Shapiro signal is least affected and gives the best reach near $10^{-2}\msun$; even so, future surveys could improve on existing NANOGrav limits by up to two orders of magnitude \cite{Cherukupalli2026}. Cherukupalli, Lee, Berghaus, and Zurek \cite{Cherukupalli2026Pairs} propose timing pulsar pairs separated by sub-parsec distances: the GWB is coherent over $d_{\rm coh}=c/(2\pi f)\simeq1$ pc at $f\simeq1.6$ nHz, so it largely cancels in the difference of the two residuals, whereas a sub-halo passing close to only one pulsar does not. For 100 pairs separated by $0.1$ pc they find an improvement of more than an order of magnitude at $\sim10^{-6}\msun$, with reach computed for masses from $10^{-11}$ to $10^{5}\msun$, and they identify globular clusters such as Terzan~5 as natural hosts, subject to separating the baryonic environment from a substructure signal. These forecasts are phrased as the minimum density of substructure for which the loudest flyby in an array reaches a signal-to-noise ratio of 4 in 90\% of realizations. Their optimism is therefore about the reach of future arrays for the abundance of substructure at masses far below the candidate's, and it is conditional on mitigating the GWB; it is not a statement about the detectability of a single $10^7\msun$ object, so it does not bear on the detection probabilities of Ref.~\cite{Wang2026} or on our corrected estimate in Sec.~\ref{sec:intro}.

\paragraph{The candidate is in the static regime.}
At $v=250~\kms$ over a $T=20$ yr baseline an object moves $vT\simeq5\times10^{-3}$ pc, far less than the $0.39$--$0.5$ kpc nominal endpoint radii of Table~\ref{tab:obs2-radii}, so the encounter is in what these papers call the static limit. The timing residual is then a power series in time. The constant and linear terms are degenerate with the pulsar's phase and frequency, and the quadratic term with its spin-down, so after the timing model is fitted only the cubic term, set by the jerk, remains \cite{Dror2019,Cherukupalli2026Pairs}:
\begin{equation}
\delta t(t)\simeq\frac{A}{r^3}\,\phi_3(t)\,\big[\hat{\mathbf v}-3(\hat{\mathbf r}\cdot\hat{\mathbf v})\hat{\mathbf r}\big]\cdot\hat{\mathbf n},\qquad
A=\frac{GMvT^3}{120\sqrt7\,c},
\label{eq:static-jerk}
\end{equation}
where $r$ is the pulsar's distance from the object, $\hat{\mathbf n}$ is the direction to the pulsar, and $\phi_3$ is the orthonormal cubic Legendre mode (peak value $\sqrt7$) \cite{Cherukupalli2026Pairs}. For $M=M_t=1.63\times10^7\msun$ and $v=250~\kms$, $A/r^3\simeq0.8$ ns at $r=0.39$ kpc and $0.4$ ns at $r=0.5$ kpc. The vector factor is at most 2, so the peak residual is at most a few ns. This is below the $50$ ns timing benchmark used in the forecasts of Ref.~\cite{Cherukupalli2026Pairs}, and realistic residuals are generally larger. A Shapiro signal is equally small: with a line-of-sight impact parameter $b$ the time-dependent part scales as $(4GM/c^3)(v_\perp T/b)^3\lesssim0.02$ ns for the impact parameters of $0.13$--$0.33$ kpc given in Appendix~\ref{app:obs2}. The gravitational effect itself is not small: an acceleration of $9\times10^{-10}~\mathrm{cm\,s^{-2}}$, the value at $0.5$ kpc for $M_t$, would produce a quadratic delay of about 6 ms over 20 yr. But a nearly constant acceleration is exactly what the timing model absorbs for a solitary pulsar. The acceleration is read out directly only through the orbital period derivative of a binary, whose intrinsic and relativistic contributions are calculable, or through an empirical spin-down calibration for solitary pulsars \cite{Donlon_SC_2025}. The acceleration search is therefore sensitive to precisely the part of the signal that timing-residual searches discard, which is why it can reach masses far above the range considered in Ref.~\cite{Dror2019}.

\paragraph{Population scale and the GWB.}
In the static limit their loudest-flyby estimate gives the median distance of the nearest object as $r_{\rm DM}=(3M\ln2/4\pi\rho_{\rm sub}N_P)^{1/3}$ for $N_P$ pulsars \cite{Cherukupalli2026Pairs}. For $M=10^7\msun$ and $\rho_{\rm sub}=f_{\rm sub}\times0.3~\mathrm{GeV\,cm^{-3}}$, this is $1.9$ kpc for $f_{\rm sub}=0.03$ and $1.3$ kpc for $f_{\rm sub}=0.1$ with one pulsar, and the line-of-sight (Shapiro) analogue is comparable or larger for this mass; at such distances Eq.~(\ref{eq:static-jerk}) is smaller than at the endpoints by a further factor of order $10^{2}$. The statement is consistent with the small prior probabilities of Sec.~\ref{sec:intro}: a $10^7\msun$ object this close to the Sun would be a rare draw, and in the PTA framework a typical member of such a population is undetectable. With the GWB included, the Doppler reach of a conventional array of 200 well-separated pulsars in the static regime is $\rho_{\rm sub}\gtrsim40~\mathrm{GeV\,cm^{-3}}$, independent of $M$ (their Eq.~S45), about two orders of magnitude above the local dark matter density. Timing residuals therefore carry essentially no information on this mass range.

The GWB is also not a limiting systematic for the acceleration observable. A background of fractional amplitude $h_c(f)$ shifts the apparent line-of-sight acceleration by roughly $2\pi f c\,h_c(f)$. For $A_{\rm GWB}=2.4\times10^{-15}$ and $\gamma=13/3$ \cite{NANOGrav2023} this is $(0.5$--$1.4)\times10^{-11}~\mathrm{cm\,s^{-2}}$ for $f$ between $1/(20~{\rm yr})$ and $1/{\rm yr}$, a factor $40$ or more below the $6\times10^{-10}~\mathrm{cm\,s^{-2}}$ residual threshold, although it should enter the covariance of a joint refit. Close-pair differencing would not help the candidate, nor is it needed. The pulsars used in the acceleration search are separated by up to $0.5$ kpc, far more than $d_{\rm coh}$, so the GWB does not cancel between them; and a smooth halo with $r_s\sim0.1$ kpc perturbs two pulsars $0.1$ pc apart identically, so differencing would remove, not preserve, its signal. In their analysis the difference channel is suppressed as $(d/r_{\rm DM})^2$ once the separation is below the size of the perturbing region, and the technique works for $r_{\rm DM}\ll d\ll d_{\rm coh}$.

\paragraph{Where the programs meet.}
Three connections are worth stating, none of which makes the candidate a PTA target. First, the two programs address different ends of the sub-halo mass function. A confirmed $10^7\msun$ object would constrain the number of massive substructures; the formation-suppression models discussed above (wave DM, WDM, DAOs) are distinguished by the abundance and structure of far lighter objects, where PTA timing is one of the few probes \cite{Dror2019,LeeMitridate2021}. One object cannot be extrapolated across some thirteen orders of magnitude in mass, but a PTA detection or bound near $10^{-6}\msun$ would give population-level information on the primordial cutoff that the candidate cannot, consistent with the conclusion of Sec.~\ref{sec:future-limits} that only a population bears on primordial suppression. Second, the compact-constituent branch of Sec.~\ref{sec:pbh} must remain consistent with global bounds on the abundance of compact objects in the mass range of Refs.~\cite{Dror2019,Ramani2020}, which SKA-era arrays could tighten if the GWB is mitigated; these bounds constrain the ambient population and not this object, and they are independent of the microlensing test of Sec.~\ref{sec:future-compact}. Third, the same pulsars supply both observables. Longer baselines improve the binary-period derivatives as $T^{-5/2}$ (Sec.~\ref{sec:future-accel}) while timing-residual searches benefit from the same baselines, so a joint treatment of the covariance, including the GWB, is natural when the catalogs grow.

\subsection{Limits of this program and the population}
\label{sec:future-limits}
Several tempting tests do not reach this object. Capture heating of substellar objects inside the halo gives equilibrium temperatures of only about 250--350 K for a brown dwarf even with geometric capture and full annihilation at densities of 0.5--2$~\msun\,\mathrm{pc^{-3}}$ and $V=250~\kms$, undetectable at 0.78 kpc. A single fast passage gives disk binaries a time-integrated exposure $\int\rho\,dt$ of order $1~\msun\,\mathrm{Myr\,pc^{-3}}$, about $1\%$ of what the smooth local halo supplies over 10 Gyr, so wide-binary disruption tests nothing here. Soliton cores and granularity at allowed boson masses lie below the scales accessible to accelerometers, and only a detected annihilation signal approaches them (Sec.~\ref{sec:future-emission}). Discreteness of a PBH-composed halo is visible only through microlensing (Sec.~\ref{sec:future-compact}) and collisional evolution (Sec.~\ref{sec:pbh}). Timing-residual searches for transiting substructure likewise do not reach an object of this mass (Sec.~\ref{sec:future-ptas}).

Resolving one profile supplies internal-structure information, not a detection of a primordial cutoff (Sec.~\ref{sec:likelihood}). A population requires further detection channels with calibrated selection functions: blind searches in the growing acceleration catalogs, the Gauss's-law method of Ref.~\cite{DonlonWidrow2025} as its sensitivity improves, unassociated extended gamma-ray sources, and, for compact constituents, microlensing. Because compact and extended objects of equal mass have different detectability in each channel, every selection function must be computed separately for each model class before it enters Eq.~\eqref{eq:expected-count}. Two channels with different selections also allow the false-positive mixture of Sec.~\ref{sec:likelihood} to be constrained from cross-detections. Counts, profiles, and velocities from such a catalog could distinguish formation suppression (WDM, wave DM, DAOs) from post-formation destruction (SIDM evaporation, decays, tides) and from PBH-enhanced initial power, subject to degeneracies with the host and primordial spectrum, and should be compared with the low-detectability alternative of Ref.~\cite{Wang2026}.

\section{Synthesis and outlook}
\label{sec:synthesis}
The candidate motivates a broad microphysical program, and the questions it raises sort naturally by the level of evidence they need: a mass perturbation, an extended bound halo, a resolved profile, or a completeness-calibrated population. Table~\ref{tab:summary} records that hierarchy, pairing each diagnostic computed here with the calculation it sets up. The numerical values in this paper are conditional calculations rather than new confidence-level exclusions, and confirmation and model comparison come before any statement that conventional CDM, canonical wave DM, a particular SIDM cross section, or a PBH fraction is favored or ruled out by this object. The corresponding opportunity is that every step along that hierarchy yields a sharper, model-specific statement about the dark sector.

\begin{table*}[t]
\caption{Summary of conditional diagnostics and the calculations they open. FDM half-power and WDM half-amplitude conventions are explicitly different.}
\label{tab:summary}
\begin{ruledtabular}
\begin{tabular}{p{0.20\textwidth}p{0.34\textwidth}p{0.38\textwidth}}
Model or mechanism & Diagnostic example & Next calculation \\
\hline
Wave-DM formation & $m_{\rm match}^{F}\sim(1.7$--$3.3)\times10^{-20}$ eV & Full transfer, progenitors, surviving population and selection \\
Wave-DM soliton & $m\gtrsim4\times10^{-21}$ eV only if $r_c<0.1$ kpc is measured and scaling applies & Direct soliton fit, envelope mass and relation scatter \\
Thermal WDM & Half-amplitude matching $m_{\rm match}^{W}\sim5.6$--7.4 keV & HMF, concentration, stripping and statistical likelihood \\
DAOs / baryon coupling & Suppression and oscillatory transfer shape & Model-specific early coupling and nonlinear calibration \\
Elastic SIDM & $t_s\sim0.16$ Gyr at $1~\mathrm{cm^2\,g^{-1}}$ & Velocity distribution, expansion/collapse branches and host interactions \\
Invisible decay & Kicks of tens of $\kms$ may alter the reference halo & Joint $(\tau_X,v_k,f_X)$ evolution and survival \\
PBHs & Shot noise and mass-dependent gravitational relaxation & Fractions, mass moments, assembly and collisional evolution \\
Prompt annihilation & Existing conditional LAT limits for $b\bar b$, $\tau^+\tau^-$ \cite{Zhu2025}; $J_{\rm ROI}=1.14\times10^{23}~\mathrm{GeV^2\,cm^{-5}}$ & Extend fixed-NFW likelihood to structural and particle models; retain residual-systematic tests \\
Radiative decay / secondaries & $D_{\rm all}\sim3.2\times10^{21}~\mathrm{GeV\,cm^{-2}}$; not bounded by the prompt-annihilation fit & New density-linear or transport maps, spectral fits and foreground treatment \\
Transiting-substructure PTA searches & Static regime: residual after timing-model fit of a few ns or less for the reference profile (Sec.~\ref{sec:future-ptas}) & Not sensitive to this mass; complementary for the low-mass end of the sub-halo mass function \\
LZ-motivated two states & $0.105~\kms$ decay kick versus $250~\kms$ downscattering at $\delta=350$ keV, $m_\chi=1$ TeV & Excited abundance, independent couplings, detector spectrum and survival \\
\end{tabular}
\end{ruledtabular}
\end{table*}

The existing prompt-annihilation search and its conditional limits belong to Zhu et al.\ \cite{Zhu2025}, and the line search to Salces P\'erez et al.\ \cite{Salces2026}; reproducing those results is a validation step, whereas varying the dynamical profile, particle model, decay channel, and transport calculation supplies the extensions identified here. The strongest immediate additions are the nonmonotonic SIDM interpretation, recoil-driven invisible decay, and the distinct PBH formation/survival channels. The LZ example makes the complementarity concrete without requiring either candidate to be real: one splitting can yield negligible decay heating but efficient downscattering ejection. The boosted-source interpretation is a separate, more model-dependent calculation, in which an annihilating sub-halo would supply a total $J$ comparable to that of the whole Galactic halo, whereas a decaying one would not compete with it. A cold microscopic candidate with a cutoff far below the accessible mass scale can remain compatible with every gravitational observation discussed here.

Each of these conditional diagnostics also defines a concrete next calculation, and together they form a coherent program. Its foundation is statistical: a joint multi-profile refit of the acceleration data, built on the common forward model of Appendix~\ref{app:obs2} and varying distances, centroid, normalization, size, and host together, combined with a selection function $S(M_t,\mathbf{x})$ calibrated by injection and recovery, turns the diagnostics of Sec.~\ref{sec:reference} into posteriors and model-specific expected counts through Eq.~\eqref{eq:expected-count}. On that footing, the formation-suppression models become precise targets. Scalar-field Boltzmann calculations with a calibrated FDM halo mass function and wave-specific tidal survival, a soliton-plus-envelope profile fitted directly to the accelerations, and a resolved, fragmentation-controlled WDM halo mass function would convert the scale matching of Secs.~\ref{sec:fdm}--\ref{sec:wdm} into likelihoods; explicit sterile-neutrino, ETHOS, and mixed cold--warm benchmarks would test how far the matched-WDM picture carries; and a completeness forecast for $10^6\msun$ objects in a future pulsar catalog would quantify the reach of the method.

Post-formation physics offers the richest new phenomenology. Forward-evolving SIDM halos through expanded-core, collapsed, and disrupted stages in a Milky Way potential, with a tidal boundary recomputed for realistic orbits and disk shocks, would predict the joint distribution of $(M_t,r_s,\rho_c,\gamma)$ and the fraction the pulsar search recovers, an observable that a measured profile could test directly. The same evolutionary machinery applies to an explicit two-state inelastic model, to a grid in $(\tau_X,v_k,f_X)$ for invisible decays, and to PBH-composed halos with extended mass functions, where point-mass, compact-cluster, and seed-plus-envelope fits to one acceleration likelihood would sharpen the compact branch. On the emission side, the Zhu et al.\ likelihood is a natural starting point: its templates serve as validation targets, and extending it to posterior-sampled $J$ and $D$ maps, velocity-dependent annihilation, density-linear decay templates, transported secondary emission matched to radio and X-ray survey beams, and a broader foreground ensemble (including Loop~I) leads toward a single joint likelihood of accelerations and photon counts with one set of structural priors. Such a likelihood could test whether morphology adds evidence for the gravitational candidate while allowing a gamma-ray residual an unrelated origin. Finally, an explicit multistate model with a computed excited abundance, and a boosted-daughter flux convolved with the LZ response, would turn the connection of Sec.~\ref{sec:lz} into a testable joint interpretation.

On the observational side, the program rests less on particle physics than on three measurements: the interior mass distribution, the velocity, and a population with known selection. Accelerations measured inside the tidal radius address the first most directly: two interior endpoints at different radii would separate the compact-object, cuspy, cored, and collapsed-halo interpretations of Table~\ref{tab:discriminants}, and the expected signals are large, with the number of interior tracers setting the pace. Lensing and accretion test the compact branches, and on the constituent branch microlensing would also measure the constituent mass and the halo's transverse velocity. The angular profile of any annihilation signal resolves the interior on scales of about 14 pc per degree, and a search for comoving stars tests whether the object is dark and, if it is not, measures its velocity. Timing-residual searches for transiting substructure \cite{Dror2019,Cherukupalli2026Pairs} probe masses lower by at least two orders of magnitude, making them a complementary window on the low-mass end of the same mass function. Soliton cores, wave granularity at allowed boson masses, and the separation of formation suppression from tidal evolution are population-level questions, which is what makes a calibrated catalog the long-term prize. A natural sequence is therefore to (1) test the single-compact-object branch with existing accretion and astrometric data and search for bound stars; (2) obtain interior accelerations and, if a compact-constituent fraction is viable, targeted microlensing monitoring; (3) compare model classes through forward evolution that includes tides and the host, with emission templates informed by the measured profile; and (4) build a population from detection channels with calibrated selection. The last step connects this single object to the primordial physics of small-scale structure.

\begin{acknowledgments}
This work was partly supported by the U.S. Department of Energy grant number DE-SC0010107 (SP). SC acknowledges support from NASA EPSCoR CAN AL-80NSSC24M0104.

\emph{AI assistance.} The authors used Claude (Anthropic) for numerical cross-checks of the quoted calculations and for LaTeX editing and drafting of parts of the text. The authors reviewed all content and take full responsibility for it.
\end{acknowledgments}

\appendix
\section{Radial geometry and the original analysis}
\label{app:obs2}
\subsection{Data provenance and radial geometry}
We use the v3 \texttt{data.csv} release linked by Donlon et al.\ \cite{Donlon_SC_2025}, from the repository \href{https://github.com/thomasdonlon/Empirical_Model_MSP_Spindown_Accels}{Empirical\_Model\_MSP\_Spindown\_Accels} (file \texttt{v3/data.csv}, SHA-256 \texttt{35a03b5d\ldots 59e02}). It contains 27 binary-derived and 26 additional spin-derived accelerations. It is used here as a geometry catalog, not asserted to be the exact selected fitting sample of Ref.~\cite{Chakrabarti2025}. The calculations of this appendix use only that file.

With catalog Galactic coordinates $(l,b)$ and curated distance $d$, we use
\begin{equation}
\mathbf x=(R_\odot-d\cos b\cos l,\ d\cos b\sin l,\ d\sin b),
\label{eq:obs2-coordinates}
\end{equation}
setting $Z_\odot=0$ explicitly. This specifies a reproducible approximate coordinate convention; the original fit's Solar height, frame, and covariance must be restored for inference. Table~\ref{tab:obs2-radii} in Sec.~\ref{sec:reference-params} gives the nearest endpoints. As a limited sensitivity check, we drew 50,000 independent positive Gaussian distances with the catalog means and errors, keeping the halo centroid fixed. The displayed 16th--84th-percentile ranges therefore describe \emph{distance-only} variations, not a joint posterior. Some other, less precisely localized sources can enter $r_t$ in these draws; the nominal exterior configuration is not asserted to hold over the full data uncertainty. The halo's quoted centroid errors are also comparable to the radial margins and cannot be propagated reliably by treating their marginals as independent.

For an additional geometric distinction, J1713+0747 has ray impact parameter $0.133$ kpc but endpoint radius $0.388$ kpc, while J1640+2224 has $0.327$ and $0.492$ kpc. Substituting the ray impact parameter for $r_i$ in the acceleration kernel would claim radial information that this observable does not contain. A joint timing-propagation analysis would require its own delay model.

\subsection{The original analysis, as specified by its authors}
The original analysis of Ref.~\cite{Chakrabarti2025} proceeded as follows; where this differs from the published text, the description given here applies. The search has three stages: (i) pairs of binary pulsars with separation $\leq0.5$ kpc and mean $|\Delta a|>6\times10^{-10}\,\mathrm{cm\,s^{-2}}$ relative to previously determined smooth Galactic models, with no significance requirement; (ii) a pair proceeds only if the mean of $a_{\rm los,Gal}/\sigma$ over its two members, derived with the binary-pulsar method, exceeds 3 (the values are 1.8 and 5.13 for J1640+2224 and J1713+0747, which we reproduce to within rounding; see below); (iii) binaries with $S/N>0.5$ and solitary pulsars with $S/N>1$ near the pair are added and the sub-halo and Galactic-potential parameters are fit simultaneously, with Bayes factors as described in Sec.~\ref{sec:intro}. The MCMC likelihood is Gaussian in $a_{\rm los,measured}$. For the binary pulsars, parallax (distance), proper motion, and the $\dot P_b^{\rm GR}$ error (from mass and eccentricity errors) are marginalized together. For the spin-derived accelerations, the errors on $\dot P^{\rm Shk}$ and $\dot P^{\rm GR}$ are taken from Ref.~\cite{Donlon_SC_2025} (the v3 catalog), parallax is marginalized, and a magnetic-braking jitter term is added because the catalog errors were found to be underestimated. The sub-halo term subtracts the Solar acceleration. The priors are uniform in $\log_{10}M_{\rm sub}$ within $\pm2$ dex of $10^7\msun$ and within 0.5 kpc in each Cartesian coordinate of the midpoint of J1640+2224/J1713+0747, and flat over two orders of magnitude for the Galactic-potential parameters. The binaries are selected within 1 kpc of that midpoint and the spin-derived accelerations within 2 kpc. The published text of the Letter differs from this specification in four places: it quotes a position prior of $<1$ kpc, a mass-prior range of two orders of magnitude, and a 1 kpc selection radius for both binary and spin-derived accelerations, and its Eq.~(1) has the opposite signs on the Shklovskii and gravitational-radiation terms. The analysis code uses the sign convention of Eq.~\eqref{eq:pdot} here. The Bayes factor is defined as in Sec.~\ref{sec:intro}; its error is taken from 1000 random subsamples of a quarter of the posterior samples. The pile-up of the $Z$ posterior against the edge of the position box was not resolved in the MCMC, and a position box of 1 kpc gave a lower significance of the mass. Using the public v3 catalog and the geometry of Eq.~\eqref{eq:obs2-coordinates}, we reproduce the stage-(ii) values as 1.80 and 5.14 (mean 3.47).

\subsection{Requirements for a joint refit}
A joint refit (Sec.~\ref{sec:synthesis}) should add the \emph{relative} Galactic acceleration to Eq.~\eqref{eq:obs2-forward} with the same coordinate and sign convention. Distances and shared calibration quantities belong in the latent parameter set and the covariance must retain shared errors: marginalizing a distance that enters both the acceleration estimator and the spatial kernel by two independent draws would double count or erase their correlation, and binary and spin accelerations from the same source are not independent if they share astrometry or calibration. The shape families to compare at a common pivot normalization $M_p=M(<r_p)$ (for instance $r_p=0.5$ kpc; a parameterization, not a measured mass) are a point mass, an NFW profile with free $r_s$, a finite Plummer core, and a soliton-plus-envelope profile \cite{Schive2014,MarshPop2015}, each with a finite tidal boundary and its normalization change integrated; freeing both $M_{\rm NFW}$ and $c$ at fixed $r_s$ is redundant. Priors should be stated with finite ranges and Jacobians (a flat prior in one mass definition is not flat in another), the null-halo model evaluated separately, and the posterior radial sensitivity compared with the prior. In the exterior configuration of Fig.~\ref{fig:obs2-identifiability} the likelihood is independent of the interior shape, so any conditional shape posterior simply follows the prior; a free-centroid fit may move endpoints inside the halo, so this local result does not replace the global comparison. The structural posterior should then be carried through calibrated stripping and evolution models with explicit infall and orbital priors, or fitted jointly as in Eq.~\eqref{eq:obs2-history}, reporting the mass definition used; a large spherical-overdensity radius obtained by formally extending the profile is not a tidal boundary. A valid posterior release needs the exact likelihood and selection, priors, chains, convergence diagnostics, and posterior predictive residuals; a joint refit additionally needs the jitter amplitude and its prior, the pre-MCMC smooth-model parameters, a treatment of the $Z$ posterior truncated by the position box, and the joint posterior samples.

\bibliography{references}
\end{document}